\documentclass{egpubl}
\usepackage{booktabs} 

\JournalSubmission    
\CGFccby

\usepackage[T1]{fontenc}
\usepackage{dfadobe}  

\usepackage{cite}  
\BibtexOrBiblatex
\electronicVersion
\PrintedOrElectronic
\ifpdf \usepackage[pdftex]{graphicx} \pdfcompresslevel=9
\else \usepackage[dvips]{graphicx} \fi

\usepackage{egweblnk}

\title[ENTIRE: Learning-based Volume Rendering Time Prediction]%
      {ENTIRE: Learning-based Volume Rendering Time Prediction}

\author[Z. Yin, H. Gadirov, J. Kosinka $\&$ S.Frey]
{\parbox{\textwidth}{
\centering Z. Yin$^{1}$, 
         H. Gadirov$^{1}$, 
         J. Kosinka$^{1}$, 
         S. Frey$^{1}$ 
        }
        \\
{\parbox{\textwidth}{\centering $^1$Univeristy of Groningen\\
}
}
}

\usepackage{mathptmx}                  

\usepackage{amsmath}
\usepackage{amssymb}
\usepackage[linesnumbered]{algorithm2e}
\usepackage{subfig}
\usepackage{subcaption}
\usepackage{float}
\usepackage{hyperref}
\usepackage[capitalize,noabbrev]{cleveref}

\crefname{figure}{Fig.}{Figs.}
\crefname{table}{Tab.}{Tabs.}
\crefname{equation}{Eq.}{Eqs.}
\crefname{section}{Sec.}{Secs.}
\crefname{algorithm}{Alg.}{Algs.}
\usepackage{tabularx}
\usepackage{multirow}
\usepackage{todonotes}
\usepackage[normalem]{ulem}
\usepackage{colortbl}
\usepackage{pgfplots} 
\usepackage{scalefnt}
\usepackage{stackengine}
\usepackage[normalem]{ulem}
\usepackage{acronym}
\usepackage{courier} 
\usepackage{amssymb}
\usepackage{algorithm2e}
\usepackage{txfonts}
\usepackage{tabularx}
\usepackage{adjustbox}
\usepackage{xcolor}
\useunder{\uline}{\ul}{}
\usepackage{ulem}
\usepackage{bibunits}
\usepackage{refcount}
\usepackage{lastpage}

\usepackage{siunitx}
\DeclareDocumentCommand{\acro}{}{ENTIRE}
\DeclareDocumentCommand{\acrovol}{}{VolumeNet}
\DeclareDocumentCommand{\acropred}{}{PredNet}
\DeclareDocumentCommand{\acrovolres}{}{LVRP}
\DeclareDocumentCommand{\acrool}{}{OLP}

\DeclareDocumentCommand{\modelsinglescattering}{}{*}

\ifdefined\dochanges
  \definecolor{changedcol}{RGB}{45, 42, 220}
\definecolor{changedcolTVCG}{RGB}{0., 0., 0.}
\else
  \definecolor{changedcol}{rgb}{0., 0., 0.}
  \definecolor{changedcolTVCG}{RGB}{0., 0., 0.}
\fi

\DeclareDocumentCommand{\changed}{m}{\textcolor{changedcol}{#1}}
\DeclareDocumentCommand{\changedFrom}{}{\color{changedcol}}
\DeclareDocumentCommand{\changedTo}{}{\color{black}}

\definecolor{softred}{rgb}{1.0, 0.7, 0.7}
\definecolor{softorange}{rgb}{1.0, 0.6, 0.2}
\definecolor{softgreen}{rgb}{0.3, 0.8, 0.5}
\definecolor{softblue}{rgb}{0.34, 0.71, 0.91}
\definecolor{softpurple}{rgb}{0.80, 0.47, 0.65}
\definecolor{softteal}{rgb}{0.4, 0.7, 0.7}
\definecolor{navyblue}{rgb}{0.2, 0.1, 0.7}
\definecolor{royalblue}{HTML}{4169E1}
\definecolor{crimson}{HTML}{DC143C}

\definecolor{cbReddishPurple}{rgb}{0.80, 0.47, 0.65}
\definecolor{bluishgreen}{rgb}{0.00, 0.62, 0.45}      
\definecolor{vermillion}{rgb}{0.84, 0.37, 0.00}
\definecolor{cbVermillion}{RGB}{213,94,0}

\pgfdeclareplotmark{rect}{
    \pgfpathrectangle{\pgfpoint{-4pt}{-2pt}}{\pgfpoint{12pt}{4pt}}
    \pgfusepath{stroke}
}
\begin{document}


\maketitle
\begin{abstract}
  We introduce \acro{}{}, a novel deep learning-based approach for fast and accurate volume rendering time prediction. 
  Predicting rendering time is inherently challenging due to its dependence on multiple factors, including volume data characteristics, image resolution, camera configuration, and transfer function settings.
  Our method addresses this by first extracting a feature vector that encodes structural volume properties relevant to rendering performance. 
  This feature vector is then integrated with additional rendering parameters, such as image resolution, camera setup, and transfer function settings, to produce the final prediction. 
  We evaluate \acro{}{} across multiple rendering frameworks (CPU- and GPU-based) and configurations (with and without single-scattering) on diverse datasets.
  The results demonstrate that our model achieves high prediction accuracy with fast inference speed \changed{and can be efficiently adapted to new scenarios by fine-tuning the pretrained model with few samples}. 
  Furthermore, we showcase \acro{}{}’s effectiveness in two case studies, where it enables dynamic parameter adaptation for stable frame rates and load balancing. 


\end{abstract}  
\section{Introduction}
\label{sec:introduction}

The high computational demand of rendering high-resolution volumes with millions or even billions of cells in volume grid structures presents challenges across various visualization use cases, from interactive exploration on workstations~\cite{beyer2015state,bruder2016real} and clusters~\cite{tkachev2017prediction} to generating image databases for in-situ visualization~\cite{ahrens2014image,bruder2022hybrid}.  
In these scenarios, accurate rendering time prediction is crucial for adjusting rendering algorithm parameters and task distribution, enabling stable interactive frame rates or balanced workloads.  
A recent study~\cite{bruder2019evaluating} identifies four key factors influencing rendering performance: hardware, rendering algorithm settings, camera parameters, and volume structure.  
Each factor can significantly impact rendering time, making it challenging to model their combined effects~\cite{bruder2019evaluating}.
For example, \cref{fig:time_vis} illustrates how rendering time varies depending on the volume, camera pose, and transfer function, highlighting the complexity of performance prediction.

Several application scenarios for rendering time prediction have been discussed in prior work, with two being particularly prominent.  
First, in interactive volume exploration, performance prediction helps estimate how to adjust sampling parameters to achieve stable rendering performance while maintaining the highest possible quality. 
A key use case is adapting rendering parameters before generating a frame to meet specific constraints, such as a target frame rate.
\changedFrom{}%
While the rendering times change smoothly between frames with user interaction such as small rotations, also abrupt changes can occur with fast camera movement or transfer function~(TF) adaptation where small changes can have a significant impact on both visual appearance and rendering performance~\cite{bruder2016real}. 
For instance, in cases where a user shifts opacity peaks in TF space to reveal different internal structures, the rendering time can change abruptly from one frame to another, e.g., due to early ray termination.
In such cases, adjusting the ray step size solely based on a window of prior frames falls short and can yield high frame latency.
\changedTo{}%
Second, in distributed rendering scenarios where \changed{a collection of tasks are assigned across different compute nodes (e.g., for creating Cinema-style databases with different views and TF settings~\cite{ahrens2014image})}, the goal is to balance the workload so that all nodes complete their tasks in approximately the same amount of time~\cite{frey_load_2011}. 
In both cases, \changed{a model that gives predictions directly based on each frame's individual configuration is essential.}
Establishing a time prediction model only needs to be done once, after which it can be reused (i)~for fluent interaction or (ii)~\changed{efficiently} distributing many rendering tasks across large clusters. 
Regarding the second case for instance, this can help to significantly reduce the time required for rendering tasks on high-performance computing systems~\cite{bruder2022hybrid}.

While there are prior works on predicting volume rendering timings~\cite{larsen2016performance, bruder_prediction-based_2017, bruder2016real, tkachev2017prediction, bruder2022hybrid}, these models generally suffer from requiring significant manual modeling effort, low flexibility for adaption to other factors (rendering methods, graphics cards) or application scenarios, and are based on assumptions on the inner workings of the rendering methods.
In particular, Larsen~\cite{larsen2016performance} developed an analytical model to evaluate rendering performance under explicit consideration of various cost factors (the parameters in the cost model require manual tuning).
Bruder \textit{et al.}~\cite{bruder2016real,bruder_prediction-based_2017} focused on modeling the performance impact of early ray termination.
Tkachev \textit{et al.}'s model predicts rendering time based on hardware configurations for distributed volume raycasting~\cite{tkachev2017prediction}. 
For load balancing in in-situ visualization settings for creating Cinema-style image data bases, Bruder et al.~\cite{bruder2022hybrid} predict the rendering time of \changed{volume} datasets via probing (completing a subset of rendering tasks) and from this calculate the mean value of selected samples' rendering times.
This yields a (predicted) average time for each individual frame task, but no detailed information per frame to allow for higher-accuracy load balancing. While considering more rendering tasks during probing increases prediction accuracy, it increases cost and reduces the amount of work that can be flexibly distributed.

\begin{figure*}[t]
    \centering
    \input{figures/time_vis}
    \captionsetup{font=footnotesize}
    \caption{
    Rendering time measurements across different volumes, camera poses, and transfer functions.  
    One representative volume was selected from each dataset used in this study.  
    For each volume, different camera orbits were applied, with the transfer function randomized at each position.  
    At the top of each subfigure, the corresponding rendered images (CUDA, with single-scattering) are shown.
    }
    \label{fig:time_vis}
    \vspace{-5pt}
\end{figure*}

We focus on the question: \emph{How to predict volume rendering time quickly and accurately?}
Among others, one influential factor impacting volume rendering performance is the volume itself, and it is of great importance for our rendering time prediction model to consider the volume's key features that contribute to its rendering time performance.
While directly considering a high-resolution volume with millions to billions of cells is comparatively time-consuming, in many typical analysis scenarios the volume data does not change on a frame-to-frame basis.
To account for this and achieve high prediction efficiency, our model splits the frame-time prediction procedure into two stages: \textbf{(i)} generation of the feature vector from the volume and \textbf{(ii)} time prediction.
In this way, once a feature representation is extracted from a volume, time predictions can be performed even more quickly for changes to camera position, transfer function, etc.
Crucially, even when considering changing volumes, our model predicts timings substantially faster than the actual rendering process.
The main contributions of our work are as follows:
\begin{itemize}
\item We present \acro{} (rEnderiNg TIme pREdiction network), a novel end-to-end model 
that decouples volume feature extraction (\acrovol{}) and rendering time prediction (\acropred{}).

\item \acro{} makes no assumptions about the underlying volume rendering method, dataset characteristics, or target hardware, enabling application to different volume-visualization scenarios.

\item \acro{}'s concatenation-based framework is designed for flexible extensibility and allows for incorporating additional rendering parameters (ray step size, lighting, sampling rates, etc.) by expanding the input vector, requiring no architectural changes.
\changed{\acro{} can be quickly adapted to new scenarios with few training data by fine-tuning a pre-trained model.} 

\item We evaluate \acro{} across various datasets and demonstrate its effectiveness on two use cases: steering toward stable frame rates and load balancing.

\end{itemize}

To the best of our knowledge, this is the first deep learning-based approach for dynamic volume rendering time prediction.

In the remainder of the paper, we will review related work (\cref{sec:RW}), discuss \acro{} and its design (\cref{sec:method}), and present our experimental setup and model evaluation (\cref{section:experiment}). 
We then consider two use cases (\cref{sec:cases}) in detail and conclude our work (\cref{sec:concl}).

\section{Related Work}\label{sec:RW}

\subsection{Volume Visualization}\label{sec:RW-VV}
There are two main approaches for classical volume visualization: indirect and direct volume rendering.
The indirect method converts volume data to an intermediate representation (e.g., an isosurface), while the direct method considers the data as a semi-transparent gel with physical properties and directly uses a 3D representation of it.
Seminal works of indirect methods extracting isosurfaces are marching cubes~\cite{lorensen1998marching} and marching tetrahedra~\cite{doi1991efficient}. 
For direct methods, the most prominently used approaches nowadays are based on raycasting~\cite{roth1982ray} for rendering and early ray termination (ERT) for acceleration. 
The rendering time of raycasting-based algorithms is highly dependent on hardware and parameter setting. This provides a good evaluation scenario for \acro{}.
Recently, works have been proposed that focus on deep learning-based volume representations and rendering, called neural rendering~\cite{tewari2020state}, which map a voxel to color and opacity using a neural network.
However, training a neural rendering model usually requires significant computational resources and large datasets. 

In the context of visualizing time-dependent volumes, several works aim to reduce costs and enhance performance.
Wang \textit{et al.}~\cite{wang2019ray} proposed to store data's depth information to recover the evolution while reducing the sampling rate. 
Flexpath~\cite{dayal2014flexpath} reduces data movements and optimizes the data placements to save transfer costs.
Frey \textit{et al.}~\cite{frey2017flow} introduced a method for time-step selection to reduce memory cost and speed up the rendering process.
Gross et al.~\cite{grosset2021lightweight} proposed a sub-sampling algorithm for time-dependent data that enables efficient local processing and focuses on optimizing computational resource usage.
Bruder et al.~\cite{bruder2022hybrid} introduced a hybrid in-situ approach for generating Cinema databases by dynamically distributing rendering tasks between simulation and visualization nodes using a simple prediction model.

\subsection{Volume Feature Extraction}\label{sec:RW-VFE}
Extracting a volume's features and representing them adequately has shown great importance in volume reconstruction and scene representation. Traditionally, a volume is represented via a large number of cells or points which explicitly store the volume and usually require a larger amount of memory.
Recent works allow a volume to be stored using a neural network, i.e., using an implicit neural representation (INR).
Typical works include Convolutional Occupancy Network~\cite{peng2020convolutional}, and Neural Radiance Fields (NeRF)~\cite{mildenhall2021nerf} as well as its variants~\cite{yu2021pixelnerf, reiser2021kilonerf}.

More recently, 3D Gaussian Splatting (3DGS)~\cite{kerbl20233d} has emerged as an efficient explicit-implicit hybrid representation for high-quality scene reconstruction and rendering, and several optimized variants have been proposed~\cite{wu2024recent, fei20243d, bauer2025gscache}.
Although these methods greatly improve rendering performance, their memory usage still scales with scene complexity and the number of Gaussian primitives.

To address this, Niemeyer \textit{et al.}~\cite{Niemeyer2019ICCV} introduced a network that learns deforming volumes by wrapping space and time. 
Weiss \textit{et al.}~\cite{weiss2022fast} proposed an efficient neural representation method by utilizing GPU tensorcores.
Tang \textit{et al.}~\cite{tang2024ecnr} presented ECNR which uses a unified space-time partitioning strategy to adaptively represent and compresses volumetric data. Still, these methods store volumes in network weights.
Gadirov \textit{et al.} introduced FLINT~\cite{gadirov2025flint} and HyperFLINT~\cite{gadirov2025hyperflint}, deep learning-based methods for volumetric flow estimation, density reconstruction, and ensemble exploration using convolutional neural networks and a hypernetwork, respectively.
Wu \textit{et al.} introduced HyperINR~\cite{wu2023hyperinr}, where a hypernetwork generates the weights of an INR network.

In \acro{}, capturing volume features via network weights would be possible, yet instead we opted to represent each volume as an explicit feature vector.
This approach allows for a more compact representation with significantly fewer elements in the feature vector compared to the number of weights stored with implicit representations.
This facilitates creating smaller, more efficient prediction networks and avoiding the dominance of volume network parameters over the other factors.

\subsection{Rendering Time Prediction}\label{sec:RW-RTP}

The prediction and modeling of algorithm performance have evolved as important research topics in scientific visualization.
With regard to predicting rendering time in volume rendering, several models that combine empirical measurements have been developed.
Sodhi \textit{et al.}~\cite{sodhi2008performance} created performance skeletons to model program execution times on the CPU.
Ipek \textit{et al.}~\cite{ipek2005approach} trained a neural network to predict the performance of large-scale application running specifically on CPUs.
These approaches are CPU architecture-specific, and thus inadequate for modeling rendering performance on GPUs.
Baghsorkhi \textit{et al.}~\cite{baghsorkhi2010adaptive} statically analyze GPU kernel code.
Zhang and Owen~\cite{zhang2011quantitative} modeled GPU performance by conducting micro-benchmarks on the target platform.
Lee \textit{et al.}~\cite{lee2007methods} proposed an early neural network-based approach that explicitly integrates statistical methods.
Wu \textit{et al.}~\cite{wu2015gpgpu} also applied machine learning to predict the scaling of power and time consumption of an application that scales with different GPU configurations.
These above-mentioned methods primarily focus on offline applications, whereas we specifically target interactive volume visualization.
Overall, there are fewer existing approaches that deal with performance prediction for real-time rendering compared to offline rendering.

Wimmer \textit{et al.}~\cite{wimmer2003rendering} proposed a rendering time prediction framework that primarily focuses on modeling the contributions of CPUs and GPUs.
Rizzi \textit{et al.}~\cite{rizzi2014performance} proposed an analytical model for GPU clusters' scaling behavior for parallel rendering that explicitly accounts for and manually models each rendering stage.
For in situ visualization, Larsen \textit{et al.}~\cite{larsen2016performance} modeled the performance of rasterization, ray tracing, and volume rendering analytically. They analyzed application performance on a single machine and applied statistical methods to calculate model weights.
After that, they further extended the model to parallel execution by utilizing a similar model for evaluating image compositing performance.
In contrast, \acro{} can flexibly deal with different rendering methods, hardware, and tasks without manual adjustment.

Bruder \textit{et al.}~\cite{bruder2016real} further incorporated the effects of acceleration approaches (early ray termination and empty space skipping) and how this affects rendering time in interactive volume visualization.
They then employed this for dynamically adjusting image resolution and load balancing. 
Tkachev \textit{et al.}~\cite{tkachev2017prediction} predicted rendering time based on different hardware configurations.
While they explicitly consider the inner workings of the (accelerated) raycasting method, their model implicitly learns the interplay of application and hardware, and users do not need to manually adapt the model to a given scenario.
Bruder \textit{et al.}~\cite{bruder2022hybrid} estimated time by rendering a random selection of images from poses of an arcball-style camera and using the arithmetic mean of the obtained times as the final time prediction.
While this is agnostic to inner workings of methods and hardware, it requires many images to be actually rendered before being able to make a prediction, and it cannot accurately predict the rendering time of individual frames.
\acro{} can adapt to complicated scenarios with heterogeneous render times more quickly and accurately (see \cref{sec:result} for a detailed comparison).




\begin{figure}[t]
    \centering
    \includegraphics[width=\columnwidth]{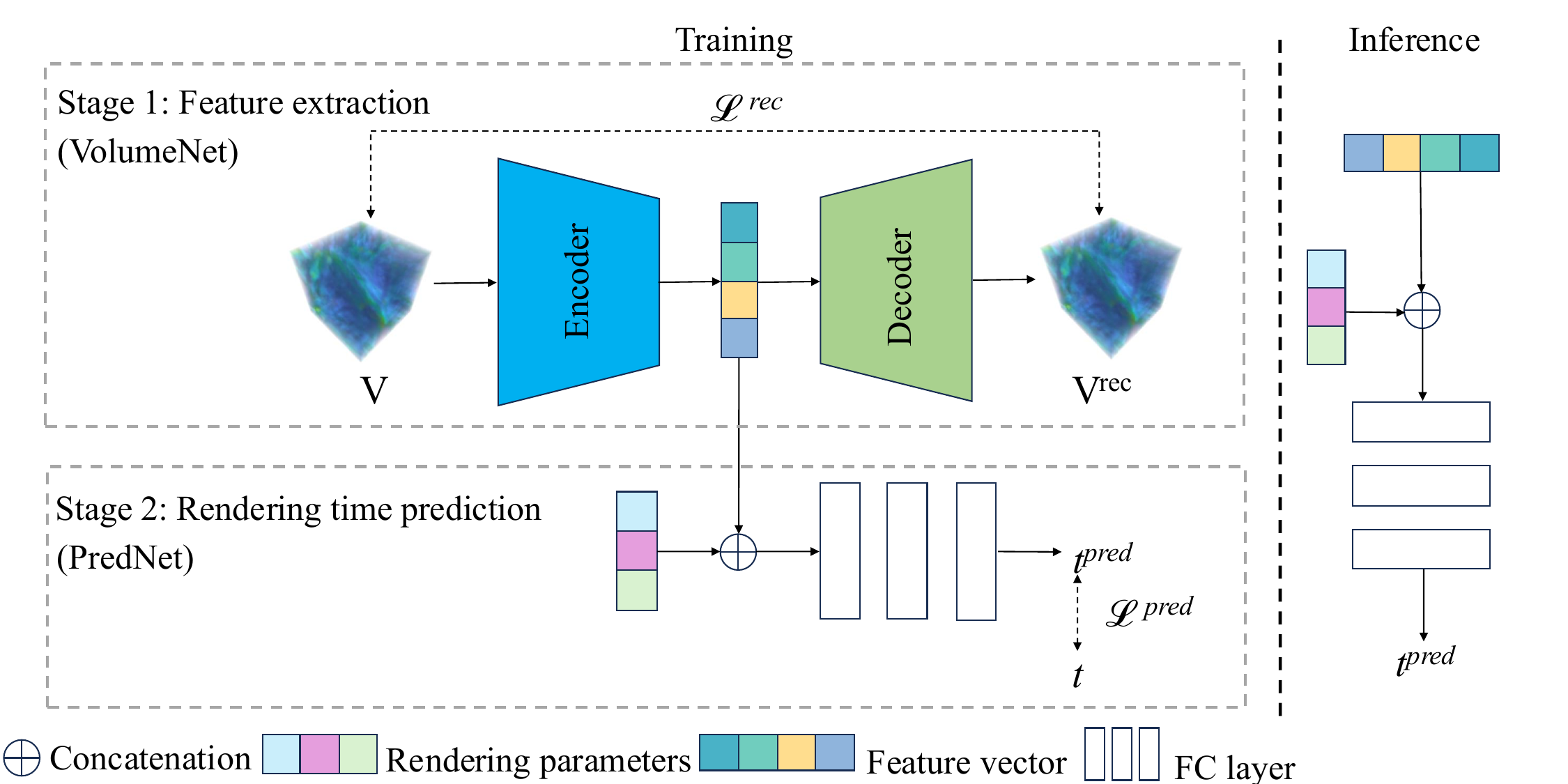}
    \captionsetup{font=footnotesize}
    \caption{Overview of the proposed prediction model. 
    Left: Model training. 
    Right: Inference. 
    $\mathcal{L}^\mathrm{rec}$ and $\mathcal{L}^\mathrm{pred}$ are reconstruction loss and prediction loss respectively, and $t^\mathrm{pred}$ is the predicted rendering time. 
    The details of \acrovol{} and \acropred{} are illustrated in \cref{fig:VolumeNet} and \cref{fig:PredNet}, respectively.
    We first train \acrovol{} for collecting feature vectors. 
    Then we train \acropred{} for rendering time prediction based on the collected feature vectors and rendering parameters.}
    \label{fig:Overview}
    \vspace{-12pt}
\end{figure}

\begin{figure}[t]
    \centering
    \includegraphics[width=\columnwidth]{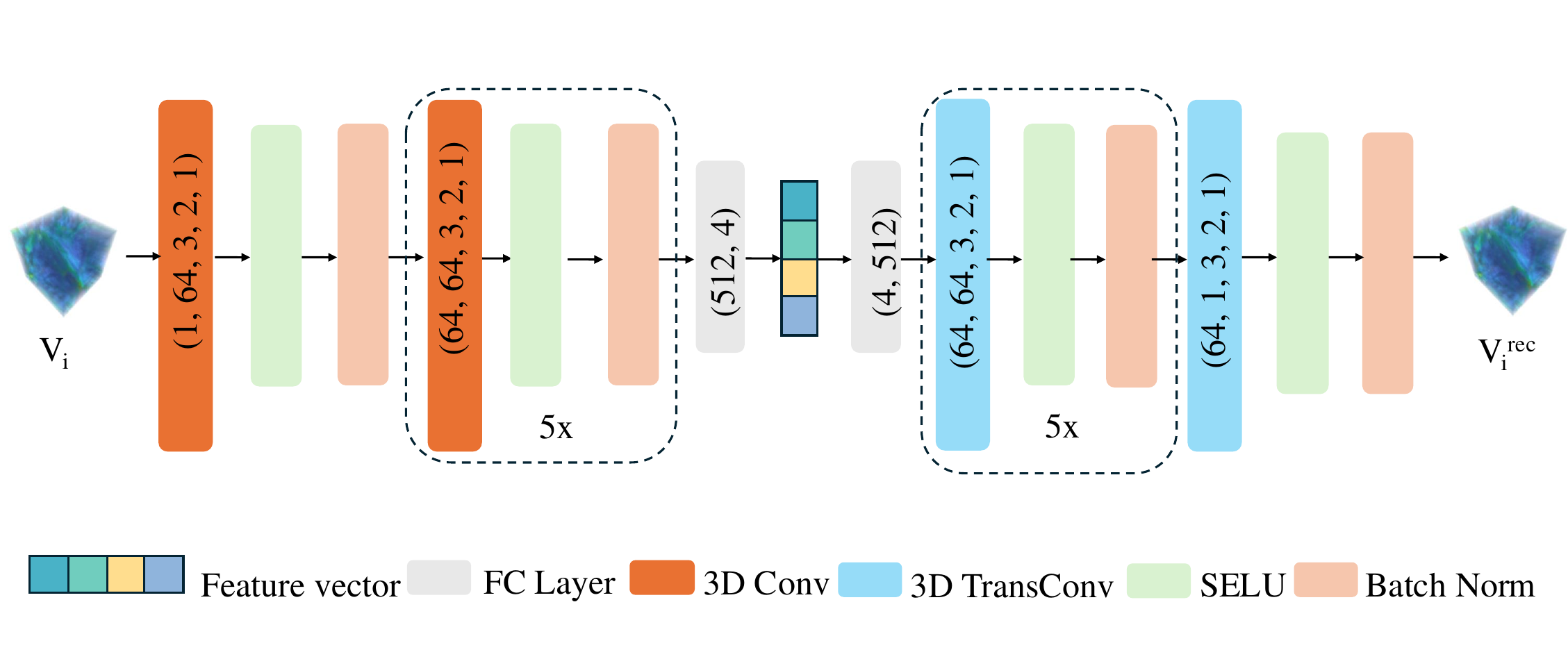}
    \vspace{-18pt}
    \captionsetup{font=footnotesize}
    \caption{
    \acrovol{}'s architecture.  
    \acrovol{} is a symmetric autoencoder with an encoder consisting of six 3D convolutional blocks followed by a fully connected (FC) layer. Each block includes a 3D convolution, batch normalization, and SELU activation~\cite{klambauer2017self}.  
    The decoder mirrors the encoder but consists of 3D transposed convolutions. The encoder compresses input data into a feature vector, while the decoder reconstructs it.  
    Layer parameters are given as (\textit{input channels, output channels, kernel size, stride, padding}) for convolutions, (\textit{input channels, output channels, kernel size, stride, padding, output padding}) for transposed convolutions, and (\textit{input channels, output channels}) for FC layers.
    We assume an input size of \(128^3\) and a feature vector of dimension 4. 
    }
    \label{fig:VolumeNet}
    \vspace{-12pt}
\end{figure}

\section{Method}
\label{sec:method}

We now describe \acro{} in detail.
\Cref{sec:method:background} reviews background on volume visualization and the challenges of time prediction.
\Cref{sec:method:overview} introduces our two-stage model, followed by detailed descriptions of its components in \cref{sec:method:volnet,sec:method:prednet}.
Finally, \cref{sec:method:jointopt} presents the loss functions.

\subsection{Background}
\label{sec:method:background}

During volume rendering, as a ray traverses the volume, the scalar value at each sample \( s \) is mapped to RGB color and opacity \( O \) via a transfer function.
These values are then accumulated to compute the final pixel color.
To optimize performance, we account for early ray termination: once the accumulated opacity reaches a predefined threshold (close to 1), further sampling along the ray is halted to save computation time.  
As a result, rendering time is influenced by the volume’s contents (specifically, voxel opacities), the applied transfer function, and the camera pose, which determines ray traversal lengths.  
Additionally, we consider single-scattering as a key factor contributing to the complexity of rendering time behavior.
Single-scattering models the interaction between light and the volume by accounting for light that is scattered once.
Incorporating this effect into rendering time prediction introduces additional light transport considerations, impacting the accumulated color and opacity along the ray.
\acro{} takes the volume, transfer function, and camera pose as inputs, enabling an end-to-end prediction of rendering time without explicitly modeling the internal computations of the rendering pipeline.

\subsection{Architecture Design}
\label{sec:method:overview}
\changed{As illustrated in \cref{fig:Overview}, \acro{} consists of two stages. Stage 1 extracts a feature vector from the volume (\acrovol{}) and Stage 2 predicts rendering time (\acropred{}).}
Considering that our target application scenarios require interactive operations, we designed \acro{}'s architecture to be as lightweight as possible to achieve a low memory footprint and fast inference speed.

\textbf{\acrovol{}.} 
First, the model transforms the considered volume into a feature representation by using an autoencoder, which consists of two components: an encoder and a decoder.
The encoder maps the data into a feature space, while the decoder reconstructs the data by remapping the feature space back to the original data space. 
In our implementation, the autoencoder leverages convolutional layers, transpose convolutional layers, and fully connected layers to enhance its feature representation capabilities (the decoder is only used for training purposes).

\begin{figure}
    \centering
    \includegraphics[width=\columnwidth]{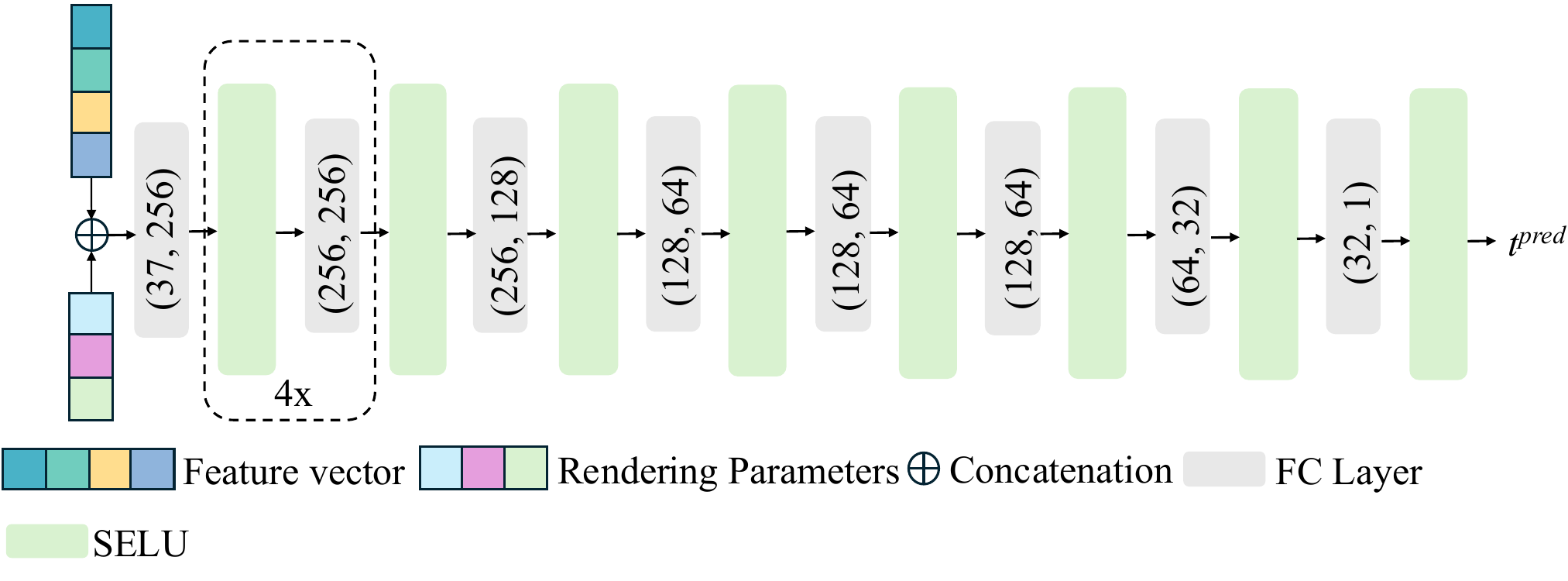}
    \captionsetup{font=footnotesize}
    \vspace{-12pt}
    \caption{\acropred{}'s architecture. 
    \acropred{} is a multi-layer perceptron with SELU activations after each fully connected layer that takes a concatenated feature vector and \changed{rendering parameters} as input, and predicts the rendering time $t^\mathrm{pred}$.
    }
    \label{fig:PredNet}
    \vspace{-12pt}
\end{figure}

\textbf{\acropred{}.} 
At the prediction stage, the model combines the learned volume feature with camera poses (comprising position and orientation) and the transfer function.
The network consists of fully-connected layers only and outputs the final rendering time prediction.
The rationale behind this two-stage design is that in many scenarios, multiple renderings of a single volume are generated by varying the camera position or transfer function.
By decoupling the model into \acrovol{} and \acropred{}, the volume is processed only once, allowing subsequent camera variations to rely solely on the compact feature representation, significantly reducing computational overhead.

\cref{fig:VolumeNet} and \cref{fig:PredNet} show the two parts of the \acro{} architecture based on our architecture selection which has three steps and is mainly focused on three architectural factors that influence the final prediction accuracy: (1)~the resolution of the volume as input to \acrovol{}, (2)~the dimensionality of the feature vector describing the volume (output of \acrovol{}, input to \acropred{}), and (3)~the number of (256, 256) layers in \acropred{}.
We aim for a model that works well across all datasets and rendering frameworks.
Accordingly, we tested models on every `rendering framework + dataset' scenario and calculated the relative deviation to the lowest prediction error.
We finally selected the model with the lowest mean relative deviation (MRD) as the final model.

\subsection{\acrovol{}: Feature Vector Extraction}
\label{sec:method:volnet}
At this stage, we extract a single feature vector from the volume using a neural network called \acrovol{} (Volume feature extraction Network, see~\cref{fig:VolumeNet}).
\acrovol{} adopts a \changed{symmetric} autoencoder architecture, where the encoder learns the volume's feature representation, and the decoder reconstructs the volume from the learned feature.
For time step $i$, where  $i=0, 1, ..., n-1$, and $X_i$ is the feature vector of volume $V_i$, 
the encoder is formulated as
\begin{equation}
    X_i = f_\mathrm{FC}^1\left(\underset{u=1...5}{\bigodot}\varphi^{u}\left(I_{\langle W_u, H_u, D_u \rangle}\right)\right),
\end{equation}
where $\varphi^{u}$ represents the $u^{th}$ convolutional layer, and $I_{\langle W_u, H_u, D_u \rangle}$ denotes the input tensor to each layer with shape $\langle W_u, H_u, D_u \rangle$, where $W$, $H$, and $D$ correspond to the width, height, and depth of the input volume, respectively. 
The function $f_\mathrm{FC}^1$ is a fully connected (FC) layer that projects the output of the convolutional layers to the feature vector $X_i$.

The decoder's objective is to recover the contents of the volume from the feature vector (for the sake of training the encoder).
After a linear projection, the feature vector is reshaped into a 3D feature map to match the input dimensions of the 3D transposed convolutional layer.
The transposed convolutional layers then produce the final predicted scalar values.

The reconstructed volume $V^\mathrm{rec}_{i}$ at time step $i$ can be described as
\begin{equation}
    V^\mathrm{rec}_{i} = \underset{v=1,...,6}{\bigodot}\sigma^{v}(J_{\langle W_v, H_v, D_v \rangle}),
\end{equation}
where $\sigma$ is the $v^{th}$ transposed convolutional layer, and $J$ is the input of each layer with shape $\langle C_v, H_v, W_v \rangle$. The input $J_{\langle W_0, H_0, D_0 \rangle}$ of the transposed convolutional layers is
\begin{equation}
    S_{\langle W_0, H_0, D_0 \rangle} = f_\mathrm{FC}^2(X_i),
    \end{equation}
where $f_\mathrm{FC}^2(\cdot)$ is the FC layer that projects $X_i$ to the dimension that $\sigma^{v}$ can process.

Note that we downsample volumes to a resolution of $128^3$ prior to passing them into PredNet for the sake of efficiency. 
\changedFrom{}%
As an autoencoder-based network, \acrovol{} is inherently lossy.
However, note that in the context of this work perfectly accurate reconstruction is not the main concern, but rather the fact that \acrovol{} produces a compact and consistent feature representation that serves as expressive input for timing prediction with \acropred{}.
\changedTo{}%

\subsection{\acropred{}: Rendering Time Prediction}
\label{sec:method:prednet}

\textbf{Camera pose.}
We employ arcball camera control in our work.
The camera pose is represented by $\eta$, a vector consisting of rotation angles $\mathbf{R} = (R_x, R_y)$ (in degrees) and translation distance $D_z$ along the z-axis for zooming.  
Throughout this work, for the sake of simplicity but without limiting the generality of our approach, the camera view is always centered on the volume.

\textbf{Transfer function representation.}
We adopt a transfer function design based on Gaussian lobes~\cite{kniss2003gaussian}, a widely used approach that is also the default in the yt visualization library~\cite{turk2010yt}.
Given a scalar volume field $s \in [0,1]$, a transfer function with $m$ Gaussian components yields respective opacity $O(s)$ as follows:
\begin{equation}
    O(s) = \sum_{i=0}^{m-1} h_i \exp\left(-\frac{(s - c_i)^2}{w_i}\right),
\end{equation}  
where $h_i$ controls the magnitude, $c_i$ determines the center, and $w_i$ defines the width of each Gaussian component. 

We represent the transfer function as the parameter vector:  
\begin{equation}
    \kappa = (c_0, w_0, h_0, ..., c_{m-1}, w_{m-1}, h_{m-1}).
\end{equation}  
This compact representation captures the essential properties of the transfer function for our model.

\textbf{Rendering time prediction.}
Our rendering time prediction network \acropred{} is depicted in \cref{fig:PredNet}.
\acropred{} (denoted as $\psi(\cdot)$) consists exclusively of fully connected layers with non-linear activation functions.
The input to the network is the concatenation of the dataset's feature vector and the camera pose. 
Then, the predicted rendering time $t_i^\mathrm{pred}$ at time step $i$ with regard to feature vector $X_i$,  camera pose $\eta$, and transfer function $\kappa$ is formulated as
\begin{equation}
    t_i^\mathrm{pred}=\psi(X_i,\eta, \kappa).
\end{equation}

\subsection{Loss Functions}
\label{sec:method:jointopt}
The volume reconstruction loss $\mathcal{L}^\mathrm{rec}$ for \acrovol{} is calculated by measuring the voxel-wise difference between the ground-truth volume $V_i$ and its reconstructed counterpart $V^\mathrm{rec}_i$:
\begin{equation}
    \mathcal{L}^\mathrm{rec}= \frac{1}{n} \sum_{i=0}^{n-1} \left|\left| V_i - V_i^\mathrm{rec} \right|\right|^{2}_{2}.
\end{equation}
Since the task of predicting rendering time is inherently a regression problem, the mean squared error (MSE) serves as the most straightforward and effective loss function.

This is also the case with the objective of minimizing the difference between predicted and actual rendering times with \acropred{}.
We quantify prediction loss via
\begin{equation}
    \mathcal{L}^\mathrm{pred}=\frac{1}{K} \sum_{i=0}^{K-1} \left|\left| t_k - t_k^\mathrm{pred} \right|\right|^{2}_{2},
\end{equation}
with $K$ and $k$ representing the  number of frames and frame index, respectively; $t_k$ and $t_k^\mathrm{pred}$ denote the ground-truth and predicted rendering time.
$\mathcal{L}^\mathrm{rec}$ and $\mathcal{L}^\mathrm{pred}$ are applied independently in the training process of Stage 1 and Stage 2, respectively (\cref{sec:method:overview}).

\section{Experiments}
\label{section:experiment}

\subsection{Experimental Setup}
\label{section:experiment:setup}
\textbf{Datasets.}
\changedFrom{}
The three datasets employed for model evaluation are listed in \cref{tab:dataset}. 
MAESTROeX~\cite{fan2019MAESTROeX} simulates low Mach number stratified flows, from which we selected the white dwarf convection problem. 
Nyx~\cite{almgren2013nyx} is a cosmological simulation of baryonic gas evolution coupled with an N-body treatment of dark matter. 
Castro~\cite{almgren2010castro} is an astrophysical hydrodynamic simulation of reacting flows, from which we selected the white dwarf merger problem.\changedTo{}
\changed{For evaluating \acro{}'s generalization capability to a different dataset, we use the CT scan of a chameleon~\cite{maisano2003}.}
To save memory space, we normalized data values in all four datasets and stored them as unsigned byte (8 bit); prior to providing volumes as input to VolumeNet, we converted  the data to  float values in $[-1, 1]$.

For MAESTROeX, we randomly divided its volumes into $80\%$ for training, $10\%$ for validation, and $10\%$ for evaluation.
We used 6 members from Nyx for training, 1 member for validation, and 1 member for evaluation.
For Castro, 8 members were used for training, 1 member for validation, and 1 member for evaluation.
\
To ensure volume diversity in our dataset, we selected every 10$^{th}$ volume from the time series and applied varying sparsity thresholds to generate multiple variants of each selected volume, creating a dataset where volumes exhibit distinct structural characteristics.
\


\begin{table}[t]
    \captionsetup{font=footnotesize}
    \caption{Dataset overview with dimensions \( M \times T \times W \times H \times D \), where \( M \) and \( T \) are the number of members and time steps, and \( W \), \( H \), \( D \) are the volume’s width, height, and depth, respectively.
    The columns ``avgRT'' reports the mean rendering time in seconds along with its standard deviation in brackets.
    The star (\modelsinglescattering) denotes the configuration where single-scattering was applied.
    \changed{The dagger ($\dagger$) indicates that the measurement was conducted on a cluster node.}
    }
    \label{tab:dataset}
    \centering
    \resizebox{\columnwidth}{!}{


\begin{tabular}{@{}ccccc@{}}
\toprule
\multirow{2}{*}{Dataset} & \multirow{2}{*}{Dimension}                     & \multicolumn{3}{c}{avgRT}                                                          \\ \cmidrule(l){3-5} 
                         &                                                & CUDA-raycaster                  & yt                & \multicolumn{1}{l}{ParaView} \\ \midrule
MAESTROeX                & 1$\times$2000$\times$512$\times$512$\times$512 & $ 0.0138 (0.0206)$              & $21.443 (33.823)$ & -                            \\ \midrule
\multirow{3}{*}{Nyx} & \multirow{3}{*}{8$\times$600$\times$512$\times$512$\times$512} & $0.0170 (0.0283)$ & $21.843 (35.365)$   & $0.0347 (0.0343)$            \\
                         &                                                & \hspace{3pt}$0.3340 (0.3565)^*$ & -                 & -                            \\
                         &                                                & $0.0054 (0.0085)\dagger$        & -                 & -                            \\ \midrule
Castro                   & 10$\times$300$\times$512$\times$512$\times$512 & $0.0124 (0.1700)$               & $21.140 (33.850)$ & -                             \\ \midrule
Chameleon                & 1$\times$1$\times$1024$\times$1024$\times$1080 & $0.0357 (0.0500)$               & -                 & -                            \\ \bottomrule
\end{tabular}%
    \vspace{-15pt}
    }
\end{table}

\textbf{Renderers.}
To evaluate \acro{}'s generalizability across GPU and CPU platforms, we utilized a CUDA volume raycaster running on the GPU~\cite{nvidia_cudasamples}, and yt~\cite{turk2010yt}(implemented in Python and executed on the CPU), which is commonly used for visualizing cosmology simulation data.
The CUDA raycaster uses early ray termination (with a threshold of 0.99), 1D transfer function lookup table, and empty space skipping for acceleration.
It employs local lighting with gradients from central differences.
Optionally, we further employ single-scattering with a termination threshold of 0.25.
The CUDA raycaster ran on an RTX 3060 GPU desktop \changed{for model evaluation}, while yt was executed on a cluster node with an AMD 7763 (32 threads).
\changedFrom{}To assess generalization to a different device, rendering times were additionally collected on a cluster node equipped with a V100 GPU.
We also ran ParaView~\cite{ParaView} on our local RTX 3060 GPU to evaluate \acro{}'s generalization ability to different renderers.
\changedTo{}

\textbf{Data collection.}
Bruder \textit{et al.}~\cite{bruder2019evaluating} highlighted that data sampled from a camera in an arcball-style orbit provides good coverage of rendering time distributions. We followed a similar strategy. Our camera followed the surfaces of several semi-spheres with various radii.
\changedFrom{}In our implementation, the renderer processes one volume at a time.
Each volume is fully loaded into memory, all its assigned frames are rendered, and only then is the next volume loaded. 
Accordingly, all reported rendering times cover the rendering process only and exclude data loading costs.
The rendered image resolution varies from $32^2$ to $1024^2$.
\changedTo{}
To ensure stable time measurements, each frame was rendered multiple times, and the median rendering time was taken as the final value. 
This is to capture performance under real-world deployment conditions (including GPU clock variations and system-level effects) rather than idealized fixed-clock scenarios. 
This approach also works consistently across all platforms (GPU and CPU) without requiring hardware-specific APIs or administrative privileges.
Each volume was rendered 100 times with random rendering parameters to collect training data.
For validation and evaluation sets, each volume was rendered 10 times using randomly chosen rendering parameters.
An overview of rendering times for each dataset is presented in \cref{tab:dataset}.
\changedFrom{}
Please note that the goal of our model evaluation is to assess \acro{}'s prediction accuracy across a broad range of rendering configurations, rather than to fit a specific user interaction scenario. 
Randomly varying parameters at each frame provides the most generic and challenging test scenario of the model's ability to predict rendering time based on each frame's individual configuration.
\changedTo{}

\textbf{Baseline.} 
We employ three different baseline models in total for comparison in different contexts.
First, we adopted the rendering time prediction method by Bruder~\textit{et al.}~\cite{bruder2022hybrid} as a baseline model for model performance evaluation. 
As discussed in \cref{sec:RW-RTP}, while many models rely on manual modeling tailored to specific rendering approaches and hardware~\cite{larsen2016performance,bruder2016real}, \acro{} is designed to be independent of these factors. 
We selected Bruder~\textit{et al.}\cite{bruder2022hybrid}'s method because it similarly generalizes across different rendering techniques and compute architectures without requiring manual adaptation~\cite{bruder2022hybrid}.  
Furthermore, their approach was developed for load balancing of a collection of rendering tasks, aligning with our goal of optimizing rendering workloads. 
Following their methodology, the baseline selects 15\% of rendering jobs from a task set and estimates the final rendering time as the arithmetic mean of the sampled times.

\
Second, \acrovolres{} (Low Volume Resolution-based Prediction) exploits the correlation between rendering times at different volume resolutions.
The method first renders volumes at low resolution ($\mathrm{64^3}$) to obtain timings, then applies a pre-computed linear scaling coefficient and a bias to predict high-resolution rendering times.
To determine the scaling coefficient and bias, we render the same scenes at both low and high resolutions with identical rendering parameters and fit a linear regression model to the time relationship.
\

Third, for ray step size control in Use Case~1~(\cref{sec:UC-step})---where we consider an interactive exploration scenario and with this have a temporal sequence of frames---we implemented an online-learning baseline model for rendering time prediction (called \acrool{}).
This model estimates the rendering time for the current frame by computing the average over the three preceding frames.
The rationale behind this approach is that rendering performance often exhibits temporal coherence, meaning that recent frames provide a strong indication of future rendering costs.
This online-learning strategy allows the system to dynamically adjust in response to variations in scene complexity and transfer function changes, making it another candidate for comparison.
\changedFrom{}%
In addition, we further implemented a PID (proportional–integral–derivative) controller as a reactive feedback-loop baseline.
The PID controller directly adjusts the ray step size based on the error between the actual and target rendering time of the previous frame.
The proportional, integral, and derivative coefficients were set empirically to 0.3, 0.05, and 0.1, respectively.
\changedTo{}

\textbf{Evaluation setup.} 
To evaluate \acro{}'s prediction accuracy, we consider the root mean square error (RMSE).
In addition to accuracy, we also measured the time \acro{} required (average time per prediction in milliseconds) during inference. 
We use the standard deviation (STD) of prediction errors to evaluate how rendering parameters affect prediction accuracy.
The reconstruction quality of \acrovol{} is evaluated in peak signal to noise ratio (PSNR) in the supplementary material.

\subsection{Evaluation of Rendering Time Prediction}\label{sec:result}
We evaluate \acro{}'s performance on MAESTROeX, Nyx, and Castro for yt.
For the CUDA raycaster, we conducted evaluations across two scenarios on Nyx: without/with single-scattering applied.
These varied scenarios were chosen to demonstrate the effectiveness of \acro{} across different rendering techniques.
\cref{tab:overall} presents prediction accuracy and runtime across different datasets and compares \acro{} to baselines.

Overall, we can see from \cref{tab:overall} that \acro{} exhibits high prediction accuracy with fast inference speed.
The standard deviation reported for avgRT in \cref{tab:dataset} indicates the high fluctuation in render times, which has already been shown by example in \cref{fig:time_vis}.
The comparison with the baseline model in \cref{tab:overall} shows that the RMSE of the rendering time prediction of \acro{} is substantially lower across all cases.
Note that the rendering time of (CPU-based) yt is three orders of magnitudes larger than with the (GPU-based) CUDA raycaster, which is also reflected accordingly in the RMSE.

\begin{table}
    \captionsetup{font=footnotesize}
    \centering
    \caption{
  Comparison of rendering time prediction accuracy in terms of RMSE (in seconds) and inference time (in milliseconds) per sample on average. 
  For \cite{bruder2022hybrid}, since its inference time is not applicable, we did not compare its inference time here.
  The star (\modelsinglescattering) denotes configurations where single-scattering was applied. The best prediction results are highlighted in bold.
  }
  \label{tab:overall}
  \resizebox{\columnwidth}{!}{%
  \begin{tabular}{@{}ccclllcllll@{}}
\toprule
\multirow{2}{*}{Framework}      & \multirow{2}{*}{Dataset}    & \multicolumn{3}{c}{\acro{}}                                                           &  & \multicolumn{2}{c}{Bruder \emph{et al.}~\@\cite{bruder2022hybrid}} &  & \multicolumn{2}{c}{\acrovolres{}}                                                     \\ \cmidrule(lr){3-5} \cmidrule(lr){7-8} \cmidrule(l){10-11} 
                                &                             & \multicolumn{3}{c}{$\mathrm{RMSE\downarrow}$ / $\mathrm{T^{pred}_{infer}\downarrow}$} &  & \multicolumn{2}{c}{$\mathrm{RMSE\downarrow}$}     &  & \multicolumn{2}{c}{$\mathrm{RMSE\downarrow}$ / $\mathrm{T^{pred}_{infer}\downarrow}$} \\ \midrule
\multirow{4}{*}{CUDA-raycaster} & MAESTROeX                   & \multicolumn{3}{c}{\textbf{0.0054 / 0.275}}                                           &  & \multicolumn{2}{c}{0.0208}                        &  &                                   \multicolumn{2}{l}{0.0474 / 27.595}                                                   \\
                                & Nyx                         & \multicolumn{3}{c}{\textbf{0.0045 / 0.274}}                                           &  & \multicolumn{2}{c}{0.0271}                        &  & \multicolumn{2}{l}{0.0190 / 8.9078}                                                    \\
                                & \hspace{3pt}Nyx$^*$ & \multicolumn{3}{c}{\textbf{0.1042 / 0.270}}                                           &  & \multicolumn{2}{c}{0.3351}                        &  & \multicolumn{2}{l}{0.3014 / 21.375}                                                                  \\
                                & Castro                      & \multicolumn{3}{c}{\textbf{0.0037 / 0.271}}                                           &  & \multicolumn{2}{c}{0.0173}                        &  & \multicolumn{2}{l}{0.0075 / 9.3684}                                                   \\ \midrule
\multirow{3}{*}{yt}             & MAESTROeX                   & \multicolumn{3}{c}{\textbf{4.337 / 0.274}}                                            &  & \multicolumn{2}{c}{34.182}                        &  & \multicolumn{2}{l}{20.63 / 2924.67}                                                                  \\
                                & Nyx                         & \multicolumn{3}{c}{\textbf{2.553 / 0.285}}                                            &  & \multicolumn{2}{c}{34.159}                        &  & \multicolumn{2}{l}{8.974 / 2611.74}                                                  \\
                                & Castro                      & \multicolumn{3}{c}{\textbf{3.541 / 0.285}}                                            &  & \multicolumn{2}{c}{34.645}                        &  & \multicolumn{2}{l}{19.35 / 2755.18}                                                                  \\ \bottomrule
\end{tabular}%
  }
\end{table}

We investigate different variants with the CUDA raycaster and the Nyx dataset.
\
Both baseline models exhibit significantly higher prediction errors compared to ENTIRE. 
\cite{bruder2022hybrid}'s method assumes homogeneous rendering time behavior, which works reasonably well when rendering parameters remain relatively stable or when jointly considering batches of rendering tasks (as in their targeted in situ visualization scenario), but the substantial variability across individual frames (see \cref{fig:time_vis}) cannot be adequately captured.
On the other hand, \acrovolres{} also shows significantly worse performance. 
Volume downsampling inherently discards fine-scale structural features that significantly influence rendering performance during raycasting.
This loss is particularly severe for our cosmology simulation datasets, leading to limited predictive capability.
\

Our yt results in \cref{tab:overall} show that with a different renderer and compute architecture, different characteristics can be observed.
For MAESTROeX, the RMSE is slightly higher than that of Nyx and Castro, not only for \acro{} but also the baseline model (although to a lesser extent; to some degree this is also the case with the CUDA raycaster).
This indicates more complex timing behavior with this data.

Notably, \acro{}'s inference both concerning \acrovol{} and \acropred{} induces only a small computational footprint: $T_\mathrm{infer}^\mathrm{vol}<\SI{10}{\milli\second}$ 
and $T_\mathrm{infer}^\mathrm{pred}<\SI{0.5}{\milli\second}$, respectively.
With this, they are substantially faster than generating an image using either renderer (avgRT in \cref{tab:dataset}).
Furthermore, an advantage of our two-stage design is that \acrovol{} needs to run only once per volume and can be reused for different camera poses.
In summary, our experiments show that \acro{} consistently achieves high accuracy and fast inference across different raycasting methods and both GPU and CPU environments.

\subsection{Impact of Rendering Parameters and Ablation Study}
\label{sec:experiment:qq}

\
To evaluate how each rendering parameter affects prediction accuracy, we designed four controlled experiments on the Nyx dataset using the CUDA raycaster.
\

\begin{figure*}[t]
    \centering
    \input{figures/nyx_results_new}
    \captionsetup{font=footnotesize}
    \caption{
    Performance of \acro{} and \acrovolres{} on different scenarios.
    For (a), (c), and (d), the frame index is indicated on $\mathrm{x}$ axis.
    For (b), $\mathrm{x}$ axis indicates the image resolution.
    }
   \label{fig:nyx_results}
   \vspace{-12pt}
\end{figure*}

\textbf{Quantitative analysis (\cref{tab:quantitative}).}
\begin{table}[t]
    \captionsetup{font=footnotesize}
    \caption{
    Quantitative comparison and ablation study on the Nyx dataset using the CUDA raycaster (no single-scattering).
    For each pair of rows of (a), models were evaluated with one target rendering parameter varied;
    for (b), all rendering parameters were varied.
    }
    \centering
    \tiny
    \captionsetup[subfloat]{position=bottom}

\subfloat[Quantitative comparison\label{tab:quantitative}]{%
    \begin{minipage}[b]{0.5\columnwidth}
        \centering
        \tiny
        \begin{tabular}{lll}
\toprule
Model                 & RMSE$\downarrow$ & STD$\downarrow$ \\ \midrule
\acro{}                & \textbf{0.0010} & \textbf{0.0001} \\
w/o feature vector    &   0.0427         & 0.0027          \\ \midrule
\acro{}                & \textbf{0.0056} & \textbf{0.0001} \\
w/o image resolution  & 0.021            & 0.0009          \\ \midrule
\acro{}                & \textbf{0.0197}  & \textbf{0.0020} \\
w/o transfer function & 0.0398            & 0.0040          \\ \midrule
\acro{}                &\textbf{0.0146}   & \textbf{0.0005} \\
w/o camera pose       & 0.0366            & 0.0028          \\ \bottomrule
\end{tabular}
    \end{minipage}
}%
\hspace{0.02\columnwidth}
\subfloat[Ablation study\label{tab:ablation}]{%
    \begin{minipage}[b]{0.4\columnwidth}
        \centering
        \tiny
        \begin{tabular}{@{}lll@{}}
\toprule
Model                 & RMSE$\downarrow$ & STD$\downarrow$ \\ \midrule
\acro{}               & \textbf{0.0045}  & \textbf{0.0003} \\
w/o feature vector    & 0.0143           & 0.0021          \\
w/o image resolution  & 0.0263           & 0.0039          \\
w/o transfer function & 0.0175           & 0.0027         \\
w/o camera pose       & 0.0098           & 0.0007          \\ \bottomrule
\end{tabular}
    \end{minipage}
}
    \vspace{-12pt}
\end{table}
Each experiment varies one rendering parameter while fixing the rest, and compares a reduced variant of \acro{} not explicitly considering this parameter against standard \acro{}.
Overall, the results demonstrate that ENTIRE consistently achieves higher accuracy than the ablated models across all parameter variations. 
%
Volume variations produce the most significant results, with prediction accuracy decreasing approximately 40$\times$ when volume feature vectors are removed.
\changedFrom{}%
This indicates that \acro{}'s autoencoder produces an expressive feature vector for each volume. 
As an end-to-end prediction model, \acro{} does not require high-fidelity volume reconstruction.
Instead, it is sufficient that each volume is mapped to a distinct representation that allows \acropred{} to differentiate rendering performance characteristics across volumes.
\changedTo{}%
As we can see from the image resolution experiments, the prediction accuracy decreased by $4\times$ when resolution information is removed.
This performance degradation confirms that image resolution is an essential factor that fundamentally determines rendering cost and cannot be omitted from accurate time prediction models.
Different transfer functions can produce vastly different rendering times for the same volume and camera configuration, depending on how they map scalar values to opacity. 
\acro{}'s learned transfer function representation successfully captures these non-linear effects, maintaining stable predictions.
Camera pose is another critical factor for rendering time prediction. 
With camera pose ablated, \acro{} showed nearly $3\times$ degradation of prediction accuracy.
Note that these experiments were designed to emphasize the impact of these parameters, and the performance impact in practical scenarios would be different (also see the ablation study below).


\textbf{Qualitative analysis (\cref{fig:nyx_results}).}
When only the volume varies, rendering times show substantial variation (\cref{fig:nyx_results:volume}).
The close alignment between predicted and ground-truth (GT) curves demonstrates that \acrovol{}'s learned feature representations effectively capture the timing-relevant properties of volumes.
When varying only the image resolution (\cref{fig:nyx_results:img_res}), ENTIRE demonstrates highly accurate predictions for low image resolution, but we observed slightly higher prediction with higher image resolution. 
Varying only the transfer function (\cref{fig:nyx_results:tf}) or camera pose (\cref{fig:nyx_results:cam_pose}) reveals more complex timing behavior.
\cref{fig:nyx_results:tf} and \cref{fig:nyx_results:cam_pose} shows significant variation in timings,reflecting the transfer function and camera pose's impact on early ray termination.
Transfer functions that produce higher opacity values lead to earlier termination and faster rendering, while lower opacity settings require longer ray traversal. 
Similarly, camera pose determines ray traversal paths through the volume, which creates complex rendering time behavior.
\acro{} accurately accounts for these variations, including sharp peaks, demonstrating its ability to model the non-linear relationship between transfer function/camera pose and rendering performance.

Across all four parameter variations, \acro{} consistently demonstrates accurate prediction with small errors ($\mathrm{<0.02s}$), validating that the model has learned meaningful relationships between each input factor and rendering time.
However, we observe a small systematic bias: \acro{} tends to underpredict when GT rendering times are comparably large($\mathrm{\approx0.15s})$.
We attribute this to imbalance in the training data, where high-cost rendering scenarios comprise only a small fraction ($\mathrm{<10\%}$) of collected samples, which further leads to the underestimation of rare peak rendering times.

\textbf{Ablation Study (\cref{tab:ablation}).}
\label{sec:experiment:ablation}
To further evaluate the individual contribution of each rendering parameter to prediction accuracy, we performed ablation experiments on the Nyx dataset using the CUDA raycaster.
Specifically, we evaluated \acro{} and ablated models with all rendering parameters varied.
The ablation study reveals that image resolution is the dominant factor influencing rendering time prediction (as indicated by significantly increased RMSE).
The impact of the remaining parameters of (volume) feature vector, transfer function and camera pose are smaller yet still substantial.
This confirms that accurate rendering time prediction requires the integration of all considered parameters. 

\changedFrom{}
\subsection{Fine-tuning}
\label{sec:experiment:fine_tuning}
While \acro{} achieves accurate rendering time prediction, collecting training data and training \acrovol{} is computationally expensive.
In practice, it would be desirable for users to adapt \acro{} to new scenarios---different renderers, devices, or datasets---with minimal data collection effort. 
In this subsection, we investigate how a pre-trained \acro{} model can achieve comparable prediction accuracy to a fully trained model while requiring significantly fewer training samples to adapt a pre-trained model to a new target scenario. 
The pre-trained model was obtained by training \acro{} (both \acrovol{} and \acropred{}) from scratch on the combined MAESTROeX, Nyx, and Castro datasets rendered via the CUDA raycaster without scattering (totally $2.4\times10^5$ training samples for \acropred{}).
We chose this scenario as it yields the greatest diversity in both volume structure and rendering time behavior.
We then fine-tuned the pre-trained \acropred{} model using varying numbers of training samples (randomly drawn from the new target scenario).
Each experiment was repeated five times and the median RMSE was taken as the final metric to account for randomness.
The results are shown in \cref{fig:fine_tune_overall}.
Note that in the following context, the sampling ratio is expressed relative to the pre-trained model's training set.

\begin{figure*}[t]
    \centering
    \captionsetup{font=footnotesize}
    \vspace{-15pt}

\definecolor{cbVermillion}{RGB}{213,94,0}
\definecolor{cbSkyBlue}{RGB}{86,180,233}

\subfloat[ParaView]{%
    \label{fig:fine_tune_overall:paraview}
    \begin{minipage}[b]{0.32\textwidth}
        \centering
        \begin{tikzpicture}
            \begin{axis}[
                xlabel={},
                ylabel={RMSE},
                xlabel style={font=\small, yshift=5pt}, 
                xticklabel style={font=\tiny}, 
                ylabel style={yshift=-0.6cm, font=\small}, 
                yticklabel style={/pgf/number format/precision=4, font=\tiny,, xshift=0.1cm},
                scaled y ticks=false,
                width=1.15\columnwidth, 
                height=4.5cm, 
                xmin=0, xmax=1, 
                ymin=0.008, ymax=0.02,
                grid=both,
                grid style={line width=.1pt, draw=gray!30, dashed},
                major grid style={line width=.2pt, draw=gray!50, dashed},
                legend pos=north east,
                legend style={fill opacity=0.0, draw=none, text opacity=1, font=\tiny},
                axis x line=bottom,
                axis y line=left,
                xticklabel={\pgfmathparse{\tick*0.45}\pgfmathprintnumber[fixed,precision=2]{\pgfmathresult}},
            ]
                \addplot[color=softteal, mark size=2pt, thick] 
                    table[x=ratio, y=exp_1_median, col sep=comma]{files/generalization/paraview/summary_exp1.txt};
                
                \addplot[color=cbVermillion, mark size=2pt, thick] 
                    table[x=ratio, y=exp_2_median, col sep=comma]{files/generalization/paraview/summary_exp2.txt};
                
                \addplot[color=cbVermillion, mark=star, mark size=4pt, only marks] coordinates {
                    (1.0, 0.0086)
                };
            \end{axis}
        \end{tikzpicture}
    \end{minipage}%
}%
\hspace{0.5em}%
%
\subfloat[V100 cluster]{%
    \label{fig:fine_tune_overall:cluster}
    \begin{minipage}[b]{0.32\textwidth}
        \centering
        \begin{tikzpicture}
            \begin{axis}[
                xlabel={},
                xlabel style={font=\small, yshift=5pt}, 
                xticklabel style={font=\tiny},
                ylabel style={yshift=-0.6cm, font=\small}, 
                scaled y ticks=false,
                yticklabel style={/pgf/number format/fixed, /pgf/number format/precision=3, font=\tiny, xshift=0.1cm},
                yticklabel={\pgfmathprintnumber[fixed,precision=3]{\tick}},
                width=1.15\columnwidth, 
                height=4.5cm, 
                xmin=0, xmax=1, 
                ymin=0.0035, ymax=0.0072,
                grid=both,
                grid style={line width=.1pt, draw=gray!30, dashed},
                major grid style={line width=.2pt, draw=gray!50, dashed},
                legend pos=north east,
                legend style={fill opacity=0.0, draw=none, text opacity=1, font=\tiny},
                axis x line=bottom,
                axis y line=left,
                xticklabel={\pgfmathparse{\tick*0.45}\pgfmathprintnumber[fixed,precision=2]{\pgfmathresult}},
            ]
                \addplot[color=softteal, mark size=2pt, thick] 
                    table[x=ratio, y=exp_1_median, col sep=comma]{files/generalization/habrok/summary_exp1.txt};
                
                \addplot[color=cbVermillion, mark size=2pt, thick] 
                    table[x=ratio, y=exp_2_median, col sep=comma]{files/generalization/habrok/summary_exp2.txt};
                
                \addplot[color=cbVermillion, mark=star, mark size=4pt, only marks] coordinates {
                    (1.0, 0.0042)
                };
            \end{axis}
        \end{tikzpicture}
    \end{minipage}%
}%
\hspace{0.5em}%
%
\subfloat[Chameleon]{%
    \label{fig:fine_tune_overall:chameleon}
    \begin{minipage}[b]{0.32\textwidth}
        \centering
        \begin{tikzpicture}
            \begin{axis}[
                xlabel={},
                xlabel style={font=\small, yshift=5pt}, 
                xticklabel style={font=\tiny},
                ylabel style={yshift=-0.6cm, font=\small}, 
                yticklabel style={/pgf/number format/precision=4, font=\tiny, xshift=0.1cm},
                scaled y ticks=false,
                width=1.15\columnwidth, 
                height=4.5cm, 
                xmin=0, xmax=1, 
                ymin=0, ymax=0.025,
                grid=both,
                grid style={line width=.1pt, draw=gray!30, dashed},
                major grid style={line width=.2pt, draw=gray!50, dashed},
                legend pos=north east,
                legend style={fill opacity=0.0, draw=none, text opacity=1, font=\tiny, yshift=0cm},
                axis x line=bottom,
                axis y line=left,
                xticklabel={\pgfmathparse{\tick*0.27}\pgfmathprintnumber[fixed,precision=2]{\pgfmathresult}},
            ]
                \addplot[color=softteal, mark size=2pt, thick] 
                    table[x=ratio, y=exp_1_median, col sep=comma]{files/generalization/chameleon/summary_exp1.txt};
                \addlegendentry{with pretrained model}
                
                \addplot[color=cbVermillion, mark size=2pt, thick] 
                    table[x=ratio, y=exp_2_median, col sep=comma]{files/generalization/chameleon/summary_exp2.txt};
                \addlegendentry{w/o pretrained model}
                
                \addplot[color=cbVermillion, mark=star, mark size=4pt, only marks] coordinates {
                    (1.0, 0.004271232831363117)
                };
                \addlegendentry{train from scratch (full)}
            \end{axis}
        \end{tikzpicture}
    \end{minipage}%
}

\vspace{-1em}
    \caption{
    \changedFrom{}
    Fine-tuning experiment on various new scenarios. The $x$-axis denotes the fraction of training samples used relative to the dataset (sampling ratio) for training the pre-trained model.
    \changedTo{}
    }
   \label{fig:fine_tune_overall}
   \vspace{-12pt}
\end{figure*}

\begin{figure}[t]
    \centering
    \captionsetup{font=footnotesize}
    \input{figures/fine_tune_qualitative}
    \caption{
    \changedFrom{}
    Qualitative comparison of fine-tuned models on different scenarios.
    Top row: ParaView.
    Middle row: V100 cluster.
    Bottom row: Chameleon.
    The fine-tuned models are indicated in {\color{softteal}soft teal}.
    The value before ``/'' denotes the ratio for ParaView and V100 cluster, and the value after means the ratio for Chameleon.
    \changedTo{}
    }
   \label{fig:fine_tune_qualitative}
   \vspace{-12pt}
\end{figure}

The pre-trained model provides a benifical initialization for fine-tuning: even with very few training samples (as few as \SI{0.45}{\percent} for ParaView and V100 cluster, and \SI{0.1}{\percent} for Chameleon), the fine-tuned \acro{} achieves acceptable prediction accuracy without significant degradation.
To match the accuracy of training from scratch on the full dataset, fine-tuning the pre-trained model with approximately \SI{11.25}{\percent} suffices for the ParaView and V100 Cluster, and \SI{0.24}{\percent} for Chameleon.
When fine-tuning to ParaView (\cref{fig:fine_tune_overall:paraview}), the pre-trained model provides relatively smaller benefits. 
This is because ParaView employs different rendering techniques from the CUDA raycaster used during pre-training, which leads to ParaView's different rendering timing behavior. 
Consequently, the pre-trained weights have limited initialization benefit.
Nevertheless, ParaView also employs a raycasting algorithm and shares common patterns with our CUDA raycaster.
Consequently, a pre-trained \acro{} fine-tuned with fewer training samples still achieves better prediction accuracy than a model trained from scratch with the same number of samples.

This benefit of the pre-trained model is most significant when fine-tuning to a new device (a cluster; see \cref{fig:fine_tune_overall:cluster}) or a new dataset (Chameleon; see \cref{fig:fine_tune_overall:chameleon}), where the underlying rendering behavior remains consistent with the pre-training configuration.
Even at zero-shot, \acro{} retains high prediction accuracy for Chameleon.
The pre-trained model is already a good starting point for this scenario: adding more training samples brings little further improvement. 
The Chameleon is acquired from CT scanning a real-world reptile and thus represents a fundamentally different modality from the astrophysical simulation datasets used in pre-training. 
Accordingly, the extracted feature vector does not fully encode Chameleon's volume information, resulting in lower reconstruction quality (approximately 15dB in PSNR). 
However, this degraded feature representation is still sufficiently expressive to yield accurate predictions accuracy for Chameleon. 
This finding further verifies our earlier discussion in \cref{sec:experiment:qq} that reconstruction quality is not the most important metric for the feature vector in our context.
What matters is that the representation is consistent across predictions for the same volume and distinguishable across different volumes, providing a stable and unique conditioning signal for \acropred{} rather than a high-quality volumetric reconstruction.
Hence, the pre-trained \acrovol{} can be applied to an ensemble or a single volume 
to obtain feature representations without retraining.
The time cost is negligible, as \acrovol{}'s inference time is under \SI{10}{\milli\second} (\cref{sec:result}).


To further validate that the fine-tuned model generalizes rather than overfits to the training data, we conducted experiments on two representative interactive visualization scenarios, where users only vary camera pose or transfer functions~(\cref{fig:fine_tune_qualitative}).
We selected models fine-tuned with \SI{0.45}{\percent}, \SI{2.25}{\percent}, and \SI{11.25}{\percent} of training samples for ParaView and V100 Cluster, and \SI{0.27}{\percent}, \SI{1.35}{\percent}, and \SI{6.75}{\percent} for Chameleon.
Even with a performance drop, fine-tuning the pre-trained model with as few as \SI{0.45}{\percent} of training samples for ParaView and Cluster, and \SI{0.27}{\percent} for Chameleon, still yields reasonable prediction accuracy, while increasing to \SI{2.25}{\percent} and \SI{1.35}{\percent} respectively does not yield substantial further improvement.


Overall, \acro{} generalizes effectively to new renderers, devices, and datasets by fine-tuning the pre-trained model with only a small number of training samples.
Users can deploy \acro{} rapidly in their target scenario with minimal data collection effort.
For quick adaptation to a new scenario, \SI{0.45}{\percent} of training samples for ParaView and V100 Cluster, and \SI{0.1}{\percent} for Chameleon, already yields acceptable performance. 
For users seeking full prediction accuracy, \SI{11.25}{\percent} and \SI{0.24}{\percent} respectively suffice.
\changedTo{}

\section{Use Cases}
\label{sec:cases}
We now discuss two use cases of \acro{}: ray step size control (\cref{sec:UC-step}) and load balancing (\cref{sec:UC-load}).
Note that in \cref{sec:UC-step} the prediction and ray step size control are conducted on-the-fly while the evaluation in \cref{sec:UC-load} considers rendering times collected a priori. Here we tested our use cases without single-scattering.

\subsection{Steering Ray Step Size for Interactive Frame Rates}\label{sec:UC-step}
Achieving interactive frame rates reliably is essential to enable fluent exploration for many visualization tasks. 
In this use case, our goal is to render frames while not exceeding a certain timing bound $t^\mathrm{target}$.
The means to achieve this is by controlling the ray step size, $\delta$, on the basis of predicted rendering time $t^\mathrm{pred}$.
For this, we experimentally determine a function $G(\cdot)$ that models the relative impact that ray step size has on rendering times.
This is done by sampling rendering times of different volumes from different poses first, and normalizing them with respect to what is achieved with reference step size $\delta^\mathrm{ref}$, yielding $t^\mathrm{norm}$.

\SetKwComment{Comment}{/* }{ */}
\RestyleAlgo{ruled}
\SetKwInput{KwData}{Input}
\SetKwInput{KwResult}{Output}
\begin{algorithm}[b]
\caption{Steering ray step size. In our evaluation, we consider a camera \emph{path} and transfer funtions $\kappa$ with a duration of $T$.}
\label{alg:control}
\KwData{\emph{path}, $\kappa$, $T$, $t^\mathrm{target}$, $G$, \acro{}, \emph{volume}}
 $\mathrm{timer.start()}$\;
 \While{$\mathrm{timer} < T$}{
  \emph{pose} $=$ \emph{path}$(\mathrm{timer})$\;
  \emph{tf} $=$ $\kappa(\mathrm{timer})$\;
  $t^\mathrm{predict} =$ \acro{}$($\emph{volume}, \emph{pose}, \emph{tf}$)$\;
  $\delta^\mathrm{adapt} = \delta^\mathrm{ref} G^{-1}\left(\frac{t^\mathrm{target}}{t^\mathrm{predict}}\right)$\;
  renderer$($\emph{volume}, \emph{pose}, \emph{tf}, $\delta^\mathrm{adapt})$\;
 }
\end{algorithm}

With this, controlling the step size for each frame works as presented in \cref{alg:control}: for each \emph{pose} along the path, obtain a time prediction from \acro{} (which can be done in less than $\SI{0.3}{\milli\second}$ as discussed above), and then using $G$ we determine what adjustment to make from $\delta^\mathrm{ref}$ to $\delta^\mathrm{adapt}$ to eventually reach the target rendering time when rendering.

\begin{figure*}[t]
    \captionsetup{font=footnotesize}
    \input{figures/use_case_1_results_new}
    \caption{\changedFrom{}%
    Steering the ray step size for interactive frame rates during an exploration session of a Nyx time step.
    (a) Top to bottom, for \acrool{}, PID, and \acro{}: 
    the measured rendering time, the adapted ray step size $\delta^\mathrm{adapt}$,
    and the measured image quality in PSNR.
    The frame index is indicated on the $x$-axis.
    \scalebox{0.5}{{\color{bluishgreen}\fboxrule=2pt\fbox{\rule{10pt}{0pt}\rule{0pt}{4pt}}}}~indicates phases with drastic rendering time changes.
    {\color{bluishgreen}$\triangle$} markers above the $x$-axis indicate the frames selected for qualitative comparison in~(b).
    (b)~Selected rendered images at frames 6, 50, and 133 (top to bottom) for GT, \acrool{}, the PID controller, and \acro{} (left to right).
    \changedTo{}%
    }
   \label{fig:control}
   \vspace{-18pt}
\end{figure*}




The experiments in this use case employ the CUDA raycaster and \changed{Nyx}.
Here, we utilized the \acro{}{} model trained with randomly varied transfer function configurations and without single-scattering (refer to line 2 in \cref{tab:overall}).
\changed{Our study depicts a scenario where the user explores a volume by adjusting the transfer function and camera pose.}
The target rendering time is set to 25~ms, corresponding to a frame rate of 40~FPS.
\changedFrom{}
For \acrool{}~(see \cref{section:experiment:setup}), we normalize each time measurement ($\mathrm{t^{adapt}}$) conducted with the respective current steps size to an estimated $\mathrm{t^{ref}}$ that reflects the reference time step before averaging, as:
\begin{equation}
\mathrm{t^{\text{ref}}} = \frac{\mathrm{t^{\text{adapt}}}}{\mathrm{G\left(\frac{\delta^\mathrm{adapt}}{\delta^\mathrm{ref}}\right)}},
\end{equation}
where $\mathrm{G\left(\frac{\delta^\mathrm{adapt}}{\delta^\mathrm{ref}}\right)}$ captures the relative rendering time change due to a changed step size.
With this, $\mathrm{t^{ref}}$ quantifies timing with respect to the reference time step just like ENTIRE's prediction does.
To evaluate the rendering quality after ray step size adjustment, we use peak signal-to-noise ratio (PSNR).
\changedTo{}

Results for ray step size steering based on this approach are shown in \cref{fig:control}.  
\acro{} achieves more stable frame rates compared to baselines.
\changedFrom{}%
During the sequences that have dramatic rendering time changes induced by the transfer function adjustment (frames 0--10, 44--53, 118--141), \acrool{} and the PID controller struggle to adapt, whereas \acro{} maintains stable predictions throughout.
When rendering time changes smoothly, e.g., the camera is moving in a smooth orbit in later phases, \acrool{} and PID even achieve slightly more stable frame rates than \acro{}.
While \acro{} occasionally produces some fluctuations due to misprediction (e.g., frames 49 and 126), it significantly outperforms both \acrool{} and the PID controller, which suffer from several practical limitations. 
First, the PID controller requires careful manual tuning of the proportional, integral, and derivative settings, and the optimal values differ across volumes, renderers, and hardware configurations. 
Second, \acrool{} and PID controller cannot respond effectively to drastic rendering time changes, as their adjustments are based on past frames that may no longer adequately reflect the current rendering cost.
Third, even under smooth rendering time changes where their adjustments are relatively accurate, being purely reactive, both \acrool{} and PID need at least one frame to `warm up', and the step size adjustment only takes effect from the second frame onward.
As a result, they always lag by at least one frame.
In contrast, \acro{} predicts rendering time before each frame is rendered and adjusts the step size proactively, without requiring any manual tuning.

Regarding the image quality, as shown in \cref{fig:control:a}\,(bottom), all three methods achieve high PSNR values throughout the sequence.
We present the rendered images from all three methods in \cref{fig:control:b}.
The differences in PSNR across the three methods are minor, and no temporal coherence artifacts were observed in the test sequence.
Hence, we argue that our method achieves stable frame rate at the cost of imperceptible degradation in image quality.

We acknowledge that \acro{} has fundamentally different cost profiles compared to \acrool{} and the PID controller: they require no training and are immediately applicable (although PID requires some parameter tuning), whereas \acro{} needs upfront investment in data collection and model training.
However, the training is a one-time investment per renderer and device configuration. 
Given interactive visualization scenarios targeting in-depth data analysis, the one-time training cost is quickly amortized.
Furthermore, as shown in ~\cref{sec:experiment:fine_tuning}, fine-tuning the pre-trained model substantially reduces this overhead when adapting to new scenarios.
\changedTo{}%

\subsection{Load Balancing}\label{sec:UC-load}

\begin{figure}
  \captionsetup{font=footnotesize}
  \includegraphics[trim=0 30 0 0, clip, width=\columnwidth]{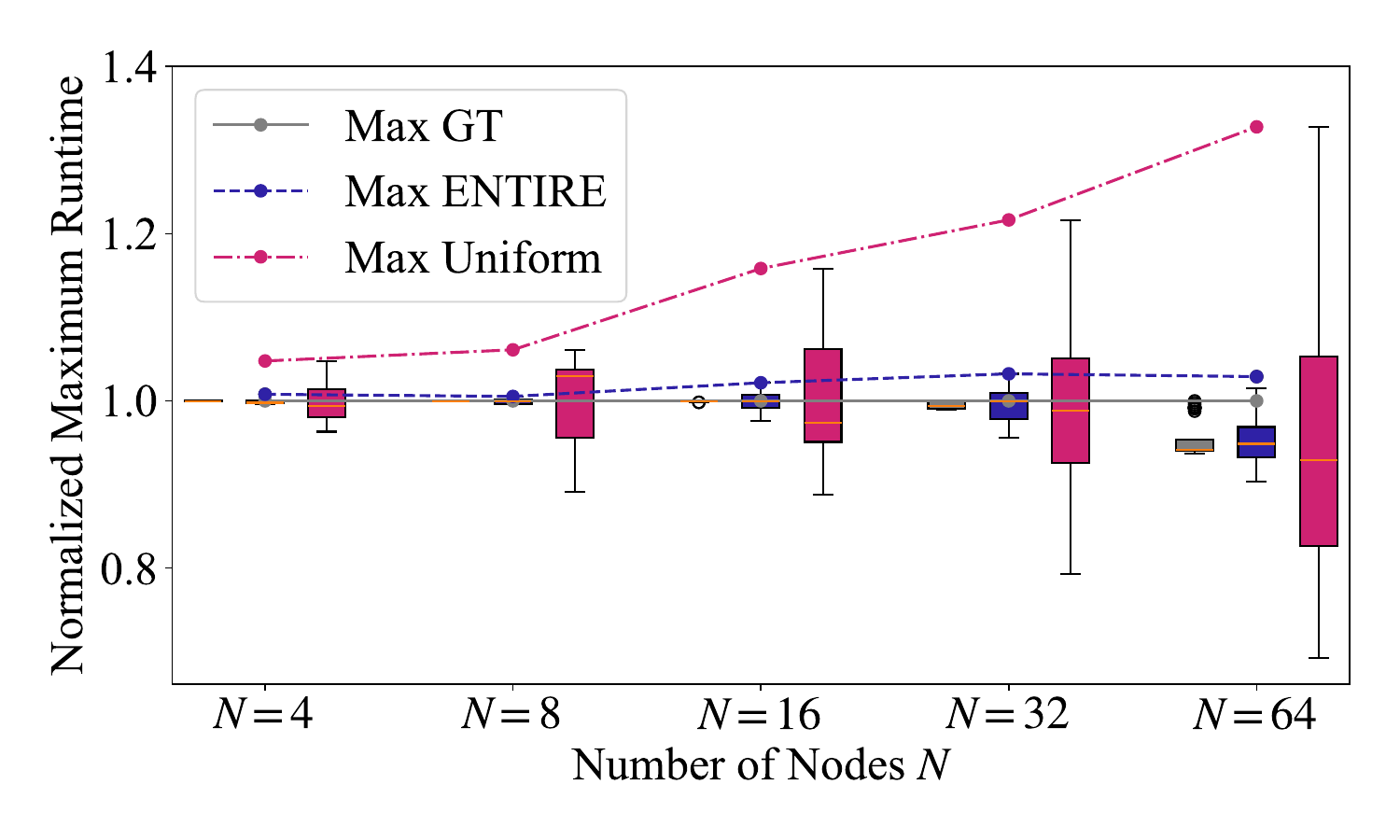}
  \caption{\changed{Load balancing use case.}
    Boxplots show the distribution of total rendering times ($\sum_{v \in A(n)} t(v)$) across nodes $n\in N$.
    Lines highlight the maximum render time that determines the runtime until completion of all tasks.
    All rendering times are normalized with respect to the GT runtime.}
  \label{fig:loadbalancing}
  \vspace{-12pt}
\end{figure}

We now present a distributed volume rendering use case involving the CUDA renderer and the Nyx dataset. In this scenario, we generate a database of images spanning six time steps, each with 64 different camera poses along an orbit, resulting in a total of \( 6 \times 64 = 384 \) visualization tasks.  
The objective is to distribute these rendering tasks across $N$ identical machines such that the maximum completion time across all machines (or nodes) \( n \in N \) is minimized. An optimal assignment should also balance the workload evenly. Mathematically, the goal is to determine an assignment \( A: N \to V \) that maps nodes \( N \) to visualization tasks \( V \), optimizing:
\begin{equation}
  \min_{A: N\to V} \left( \max_{n \in N} \left(\sum_{v \in A(n)} t(v)\right) \right),
  \label{eq:loadbalancing}
\end{equation}
where \( t: v\to\mathbb{R} \) provides an estimate of the anticipated rendering time for each task \( v \in V \).  

This problem is NP-hard, even under the assumption of identical machines. To address it, we employ the Longest Processing Time First (LPT) algorithm, a well-established greedy approach with a tight error bound~\cite{graham:69}. LPT assigns rendering tasks in descending order of cost—here, based on predicted rendering times \( t(\cdot) \),—iteratively distributing them to nodes in a way that minimizes the maximum load after each step.

We compare three approaches for estimating the rendering time \( t(\cdot) \) used in \cref{eq:loadbalancing}: \textbf{GT}: Measured ground truth rendering times, assuming perfect predictions; \textbf{\acro{}}: Rendering times predicted by \acro{}; \textbf{Uniform}: All tasks are assumed to have the same (uniform) rendering time.
With the Uniform approach, LPT assigns tasks in a random order, resulting in a balanced task count but potentially suboptimal scheduling.  

\cref{fig:loadbalancing} presents the results for varying numbers of nodes $N$, where rendering times are normalized relative to the slowest node in the GT case (i.e., the theoretical optimum from \cref{eq:loadbalancing}). Each node count $N$ is associated with three boxplots representing GT, \acro{}, and Uniform, showing the distribution of total rendering times \( \sum_{v \in A(n)} t(v) \) across nodes \( n \in N \).  
The boxplots reveal that as $N$ increases, load imbalance grows for both \acro{} and Uniform relative to the GT baseline. 
The three curves overlaid on the boxplots in \cref{fig:loadbalancing} represent the maximum render time for each method, directly reflecting the objective value in \cref{eq:loadbalancing}. 

The relative difference from GT increases with $N$ for both \acro{} and Uniform. 
While \acro{} incurs less than \( 4\% \) of additional cost compared to GT even for large $N$ due to its high prediction accuracy, Uniform suffers a significantly higher overhead, nearing \( 20\% \).  
Overall, these results highlight the effectiveness of \acro{} in large-scale volume visualization, closely approximating the theoretical optimum while significantly reducing computational overhead.

\section{Discussion}\label{sec:discussion}


\textbf{Multi-dimensional performance analysis.}
The evaluation in \cref{sec:result} shows that various factors influence not only the rendering time but also the prediction quality as they introduce complexity in timing behavior. 
Additionally, the comparison across different datasets, such as the Nyx dataset and the MAESTROeX dataset, highlights that the rendering method and dataset-specific characteristics play a significant role in rendering performance.
The results obtained using the CUDA raycaster and yt show that while comparable prediction accuracy can generally be achieved, there are noticeable differences, especially in the case of the MAESTROeX dataset that require further investigation.
An extended, adaptive performance assessment effort could help to account for this in future work.
Specifically, we plan to incorporate \acro{} with \cite{bruder2019evaluating}'s method for performance assessment.

\textbf{Generalization.}
\cref{sec:result} has demonstrated that \acro{} can generalize without specific architecture adjustment across various configurations: different rendering codes and hardware (CUDA-raycaster on the GPU and yt on the CPU), raycasting variants (with/without single scattering), different transfer function configurations, and datasets.
\changed{When the user is working with different renderers, devices, or datasets, they can fine-tune the pre-trained \acro{} model with relatively few samples to achieve reasonable prediction accuracy.}
Furthermore, \acro{}'s concatenation-based framework enables flexible extension of the model to the target application scenario by simply adding required input parameters.


\textbf{Training efficiency.}
Although \acro{} achieves accurate rendering time prediction on-the-fly, the collection of training data remains a computationally expensive process. 
Even with acceleration techniques such as early ray termination,  and empty space skipping, collecting rendering times for Nyx's training set required over 18 hours on the GPU. 
This represents a practical obstacle for users seeking to apply \acro{} to new datasets immediately.
Inspired by ~\cite{herveau2021analysis}, a promising direction for future work is to train ENTIRE via online learning.
Specifically, \acro{} is trained incrementally as users render frames during exploration. 
Once \acro{} reaches sufficient prediction accuracy (e.g., below a target threshold), the training process terminates.
This approach combines data collection and model training, eliminating the need for dedicated upfront data collection and enabling users to apply \acro{} to new datasets immediately.
Currently, \acrovol{} and \acropred{} are trained separately. 
We aim to investigate joint training strategies, which could potentially improve performance by allowing the two components to learn more cohesive representations \changed{that are specifically tuned toward rendering time prediction}. 
\changedFrom{}%
Such an architecture could potentially yield more task-relevant feature representations compared to our current decoupled design.
\changedTo{}%
However, this also introduces challenges, particularly increased memory consumption and training complexity, which we plan to address.


\textbf{Ray Step Size Adaptation (Use Case 1).}
As shown in Sec. 5.1, the rendering times achieved by \acro{} are generally slightly slower than the target times due to the simple approach how we adapt the ray step size with $G$.
A more general and direct solution is to extend \acro{} to additionally take the ray step size $\delta$ as an input.
While this increases the model complexity and the amount and variation of training data required, this would also be the most flexible method that could further capture complex interdependencies between $\delta$ and other influence factors \cite{bruder2019evaluating}.

\textbf{Load balancing (Use case 2).}
\cref{sec:UC-load} shows the effectiveness of \acro{}'s timing predictions in load-balancing in a distributed rendering scenario by demonstrating only small inefficiency in comparison to a hypothetical perfect prediction model (GT).
To be able to make this comparison, we relied on measurements conducted a priori. In future work we plan to implement \acro{} load balancing in existing distributed rendering frameworks. 
In doing so, we further aim to extend the investigation of \acro{}'s utility for different use cases in parallel volume rendering~\cite{molnar:94}---especially sort-last approaches---and also evaluate them with practical experiments on clusters and large-scale compute infrastructures.
Related to our targeted generalization efforts described above, we also aim to experiment with applying \acro{} to other timing predictions tasks, e.g., with the compositing step required for sort-last methods.

\textbf{Future work.}
We plan to explore the application and extension of \acro{} to predict rendering quality/error alongside rendering time for given renderer parameters.
With both predictions, the system could select parameter combinations (e.g., ray step size and image resolution) that lie on the Pareto frontier, achieving optimal quality within a specified time budget.
\changedFrom{}%
Furthermore, the current evaluation assumes volumes are fully pre-loaded into memory and data loading costs are excluded from all reported rendering times.
Extending \acro{} to account for data loading and caching effects is a promising direction for future work.
\changedTo{}%

\section{Conclusion}\label{sec:concl}
We introduced \acro{}, a novel ML-based approach for predicting volume rendering time without requiring manual adjustments for specific rendering methods, hardware, or datasets.
Our approach follows a two-stage design: (1) encoding the volume into a feature vector (\acrovol{}), and (2) leveraging this representation, along with rendering parameters,
to predict rendering time (\acropred{}).
We demonstrated that \acro{} achieves both high accuracy and fast inference across diverse datasets, computing environments, and rendering techniques (even in the presence of complex interactions in single-scattering scenarios).
\changed{We also showed that a user can fine-tune a pre-trained \acro{} model with relatively few training samples for fast adaptation to new scenarios.}
Furthermore, we demonstrated \acro{}'s practical utility in two key scenarios: (1)~dynamically adapting rendering parameters to optimize computational cost while maintaining performance targets, and (2)~balancing workloads in distributed rendering setups to ensure efficient resource utilization.

\bibliographystyle{eg-alpha-doi} 
\bibliography{template}       


\newpage
\clearpage

\section*{Supplementary Material}\label{sec:supplement}
\renewcommand{\thesubsection}{\Alph{subsection}} 

\subsection{Implementation}\label{sec:supp:implemenation}
\acro{} is implemented in PyTorch.
We employed the Adam optimizer~\cite{KingBa15} and trained our model using the cosine annealing learning rate decay strategy~\cite{loshchilov2017sgdr}. Adam's momentum settings are $\beta_1=0.9$, $\beta_2=0.999$.
To prevent overfitting, early stopping was applied to halt the training process once the validation loss ceased to decrease. 
Additionally, we applied gradient clipping with a maximum norm of 1.0 to avoid gradient explosion.
\acrovol{} was trained for \changed{200} epochs and evaluated on an Nvidia A100 GPU, with batch size of 16 and decreasing learning rate in range $[10^{-3}, 10^{-5}]$.
\acropred{} was trained for 200 epochs and tested on an Nvidia RTX 3060 GPU. 
We set \acropred{}'s decreasing learning rate to $[10^{-4}, 10^{-6}]$.
Training \acrovol{} required approximately 2.5 hours, while it only took up to 5 minutes for \acropred{}.

\subsection{Scientific Dataset}\label{sec:supp:scientific_dataset}
Examples from the employed datasets can be found in \cref{fig:data_evolve}.
For MAESTROeX, the volume exhibits significant changes during the early stages, then stabilizes in the middle and late phases.
Nyx shows relatively consistent evolution throughout all time steps.
In Castro, sharp changes occur initially, particularly during the merger of two white dwarfs; then the volume enters an oscillatory phase.
We provide three videos for each dataset to illustrate their evolution over time.
\cref{fig:dataset} presents the datasets used in this study, rendered by CUDA raycaster and yt~\cite{turk2010yt}.

\begin{figure*}[t]
    \centering
    \input{figures/dataset_evolve}
    \caption{
    Examples for MAESTROeX (top row), Nyx (middle row), and Castro (bottom row).
    }
   \label{fig:data_evolve}
\end{figure*}

\begin{figure}[t]
    \centering
    \input{figures/dataset}
    \caption{Example renderings with the CUDA raycaster (top) and yt (bottom) of datasets employed in this paper. 
    }
    \label{fig:dataset}
    \vspace{-12pt}
\end{figure}

\subsection{Architecture Selection} \label{sec:supp:arch_selec}
Here, we describe how we selected the final \acro{} architecture.
To develop a model that generalizes effectively across all three datasets, we trained \acro{} on the combined data from MAESTROeX, Nyx, and Castro. 
The architecture selection followed a two-stage process: First, we evaluated \acrovol{} architectures by comparing volume reconstruction quality (measured by PSNR) and the dimension of feature vectors for the best rendering time prediction accuracy.
Second, we determined the optimal number of fully-connected layers for \acropred{} through the evaluation of prediction performance.
Below, we refer to architecture variants in a short-had form, such as $128^3$F$16\rightarrow5$C$256$, where:
\begin{itemize}
\item[F] is used to indicate the resolution of the input volume and the dimensionality of  the volume feature vector.
For instance, $128^3$F$16$ indicates a volume of dimension $128^3$ encoded into a feature vector of length $16$.

\item[C] depicts fully-connected layers. It is preceded by the number of layers, and followed by the number of channels of the first layer; the number of channels repeats for each following layer.
For instance, $16$C$256$ denotes $16$ repeated fully-connected layers, with the input and output size of $256$. 
Note that \acropred{}'s rest layers are not explicitly represented in this notation.
Each layer's dimension is halved with respect to its previous layer. 

\item[$\rightarrow$] \hspace{0.1mm} denotes the transition between a feature vector and fully-connected layers.
\end{itemize}

Accordingly, $128^3$F$16\rightarrow5$C$256$ means that \acrovol{} considers volumes with a resolution of  $128^3$ and the output feature vector's dimension is $16$.
The corresponding \acropred{} has $5$ fully-connected layers as described above.

\textbf{(1) Volume resolution.}
We tested with volume resolutions $64^3$ and $128^3$.
Training \acrovol{} on volumes with higher resolutions took more than 15 hours, which we consider to be impracticable in our target application scenarios.
Note that all volumes for training and testing here were downsampled from the original full volumes, and that this downsampling only applies to \acrovol{} input (i.e., in order to obtain the feature vector describing the volume). 
The rendering time measurements used to train \acropred{} are captured with the original volume size.
We experimented with five different feature vector sizes $\{16, 32, 64, 128, 256\}$ and four $C256$ layers.

For \acrovol{}, increasing the input volume resolution consistently led to improved volume reconstruction quality.
Interestingly, a higher input volume resolution to \acrovol{} does not necessarily lead to a higher prediction accuracy.
\cref{tab:feature_select_1} indicates for CUDA raycaster and yt that \acro{} showed its best performance with $128^3$.
From \cref{tab:feature_select_1}, we observe that prediction accuracy is jointly influenced by both input volume resolution and feature vector dimensionality.
When the input resolution is high but the feature vector dimension is low, the resulting feature vector may fail to capture sufficient structural information, thus reducing prediction accuracy.
Moreover, higher input resolution significantly increases training cost—up to a $5\times$ increase for $128^3$ volumes.
\begin{table*}[t]
\Large
\centering
  \caption{PSNR and inference time ($\mathrm{T_{infer}^{vol}}$) of \acrovol{} and Prediction RMSE values of \acropred{}.
Best prediction results are highlighted in bold.}
  \label{tab:feature_select_1}
  \setlength{\tabcolsep}{3pt}
  \resizebox{\textwidth}{!}{%
  \begin{tabular}{@{}cccc|ccc|ccc|ccc|ccc@{}}
\toprule
\multirow{3}{*}{Res} &
  \multicolumn{3}{c|}{F=16} &
  \multicolumn{3}{c|}{F=32} &
  \multicolumn{3}{c|}{F=64} &
  \multicolumn{3}{c|}{F=128} &
  \multicolumn{3}{c}{F=256} \\ \cmidrule(l){2-16} 
 &
  \multirow{2}{*}{$\mathrm{PSNR\uparrow}$/$\mathrm{T^{vol}_{infer}}\downarrow$} &
  \multicolumn{2}{c|}{$\mathrm{RMSE\downarrow}$} &
  \multirow{2}{*}{$\mathrm{PSNR\uparrow}$/$\mathrm{T^{vol}_{infer}}\downarrow$} &
  \multicolumn{2}{c|}{$\mathrm{RMSE\downarrow}$} &
  \multirow{2}{*}{$\mathrm{PSNR\uparrow}$/$\mathrm{T^{vol}_{infer}}\downarrow$} &
  \multicolumn{2}{c|}{$\mathrm{RMSE\downarrow}$} &
  \multirow{2}{*}{$\mathrm{PSNR\uparrow}$/$\mathrm{T^{vol}_{infer}}\downarrow$} &
  \multicolumn{2}{c|}{$\mathrm{RMSE\downarrow}$} &
  \multirow{2}{*}{$\mathrm{PSNR\uparrow}$} &
  \multicolumn{2}{c}{$\mathrm{RMSE\downarrow}$} \\ \cmidrule(lr){3-4} \cmidrule(lr){6-7} \cmidrule(lr){9-10} \cmidrule(lr){12-13} \cmidrule(l){15-16} 
 &
   &
  CUDA &
  yt &
   &
  CUDA &
  yt &
   &
  CUDA &
  yt &
   &
  CUDA &
  yt &
   &
  CUDA &
  yt \\ \midrule
64x64x64 &
  27.85/1.994 &
  0.0080 &
  3.879 &
  28.07/1.987 &
  0.0081 &
  \textbf{3.823} &
  28.37/1.835 &
  0.0083 &
  3.885 &
  \textbf{28.44}/1.983 &
  00086 &
  3.911 &
  28.40/1.880 &
  0.0088 &
  3.928 \\
128x128x128 &
  27.28/9.733 &
  \textbf{0.0071} &
  3.853 &
  27.42/5.769 &
  0.0080 &
  3.868 &
  27.62/5.786 &
  0.0078 &
  3.931 &
  27.74/6.410 &
  0.0084 &
  3.922 &
  27.75/5.710 &
  0.0086 &
  3.890 \\ \bottomrule
\end{tabular}
  }
\end{table*}

\textbf{(2) Feature vector dimension.}
From \cref{tab:feature_select_1}, we see that for the CUDA raycaster, \acro{}'s best volume resolution and feature vector dimension combinations are $128^3$F$16$.
For yt, the best combinations is $64^3$F$32$.

The final prediction accuracy decreased as $F$ grew from 16 for CUDA-raycaster and 32 for yt.
To further explore the feature vector dimension's effect on prediction accuracy, we took the models from \cref{tab:feature_select_1} which have the best performance and used them as the baseline model. 
If $\mathrm{F}$ in the baseline model is 16, then we decreased it until \acro{}'s prediction accuracy showed a downward trend. 
The results are presented in \cref{tab:feature_select_2}.
Note that, as the optimal configuration of volume resolution and feature vector dimension for yt has already been identified as $64^3$F$32$, we did not include further experiments for this setting in \cref{tab:feature_select_2}.
Surprisingly, reducing feature vector dimensionality yielded higher prediction accuracy. 
We decreased feature dimension from 16 to 2, finding that each reduction improved performance.
Although further reduction to $\mathrm{F=1}$ was theoretically possible, a single scalar cannot adequately represent the complex structural variations across volumes, so we set $\mathrm{F=2}$ as the minimum viable dimension.

Overall, the best input volume size and feature vector dimension for the CUDA raycaster are $128^3$F$4$, $64^3$F$16$, and $128^3$F$16$ for MAESTROeX, Nyx, and Castro.
For yt, the best input volume size and feature vector dimension are $64^3$F$32$.

\begin{figure}[t]
    \centering
    \captionsetup{font=footnotesize}
    \includegraphics[width=\columnwidth]{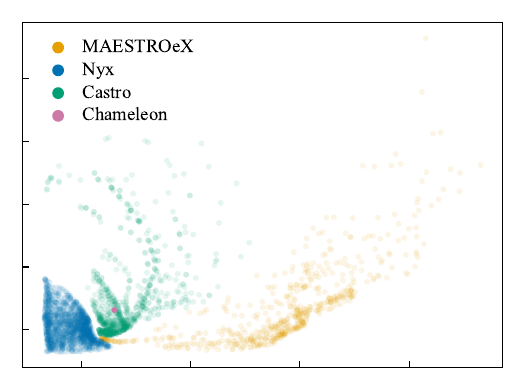}
    \caption{
    \changedFrom{}
    Visualizations of feature vectors for all training datasets and Chameleon. 
    Each point represents a volume, with opacity indicating the time step of the corresponding simulation member.
    \changedTo{}
    }
   \label{fig:umap_dataset}
   \vspace{-12pt}
\end{figure}

\begin{figure}[t]
    \centering
    \subfloat[MAESTROeX]{
      \begin{minipage}[b]{0.225\columnwidth}
      \centering
        \includegraphics[width=\columnwidth]{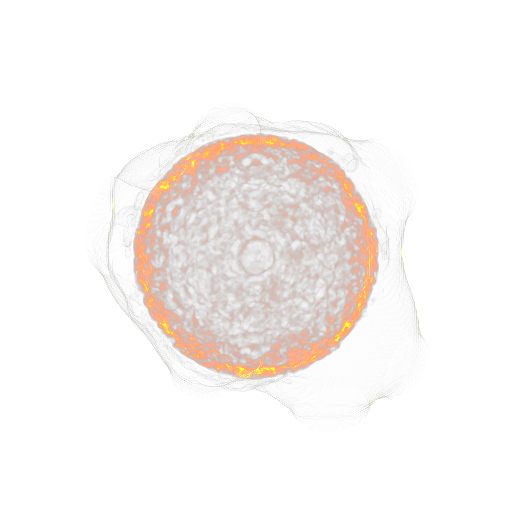} \\
        \includegraphics[width=\columnwidth]{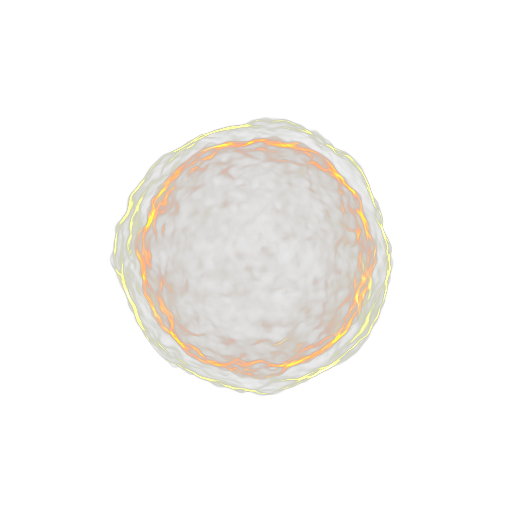} \\
      \end{minipage}
      }
    \subfloat[Nyx]{
      \begin{minipage}[b]{0.225\columnwidth}
      \centering
        \includegraphics[width=0.9\columnwidth]{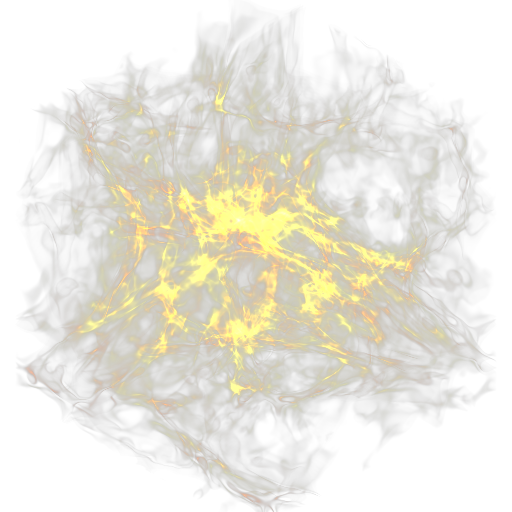} \\
        \includegraphics[width=0.9\columnwidth]{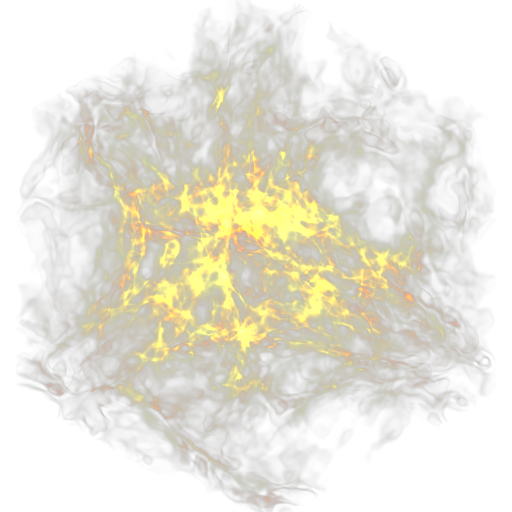} \\
      \end{minipage}
      }
    \subfloat[Castro]{
      \begin{minipage}[b]{0.225\columnwidth}
      \centering
        \includegraphics[width=\columnwidth]{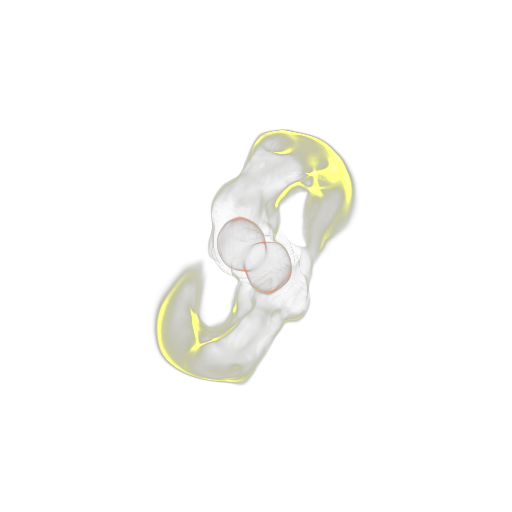} \\
        \includegraphics[width=\columnwidth]{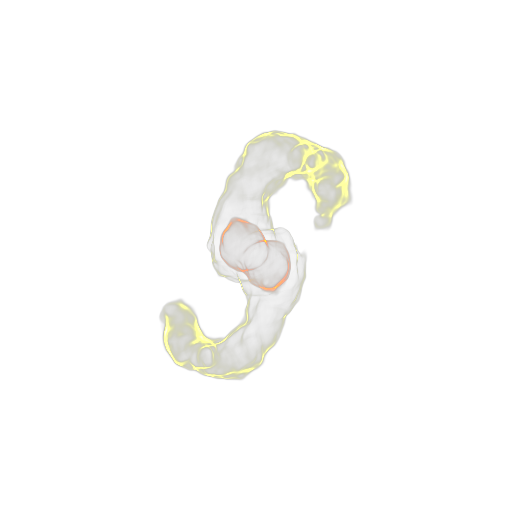} \\
      \end{minipage}
      }
    \subfloat[\changed{Chameleon}]{
    \label{fig:volunenet_reconstruction:chameleon}
      \begin{minipage}[b]{0.225\columnwidth}
      \centering
        \includegraphics[width=\columnwidth]{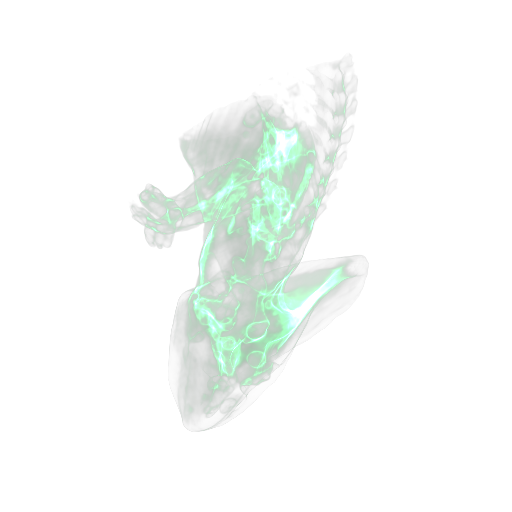} \\
        \includegraphics[width=\columnwidth]{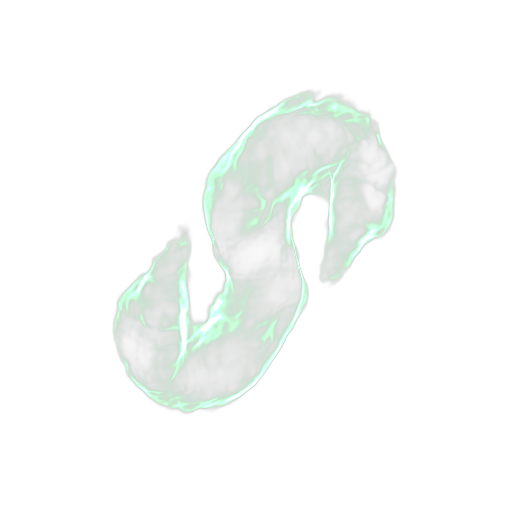} \\
      \end{minipage}
      }
    \caption{Comparison of GT (top row) and \acrovol{}'s reconstruction (bottom row).
    }
    \label{fig:volunenet_reconstruction}
\end{figure}

When the feature vector dimension $F$ becomes very high, it may include volume features that are negligible for relevance for rendering time prediction. 
This can introduce noise or shift the model’s focus away from \changed{other input} features that are more predictive, such as camera pose or \changed{transfer function}. 
Consequently, the model's ability to generalize across varying \changed{rendering} configurations may diminish, leading to reduced prediction accuracy.
\changedFrom{}This suggests that it is not simply the quality of the reconstructed volume, but the distinctiveness and size of the extracted features that determine overall performance.
We visualized the feature vectors of the pre-training datasets and Chameleon in~\cref{fig:umap_dataset}.
A qualitative comparison of \acrovol{} is shown in \cref{fig:volunenet_reconstruction}.
\acrovol{} produces well-separated clusters for the three training datasets, with minor overlap between Castro and MAESTROeX.
Nevertheless, Chameleon's feature vector falls within the Castro cluster in the feature space and \acro{} reconstructs Chameleon with a Castro-like appearance. 
As Chameleon was never seen during training and differs from the simulation datasets, \acrovol{} maps it to the nearest cluster in the learned latent space, which happens to be Castro.
This is consistent with the renderings of the two datasets (\cref{fig:volunenet_reconstruction:chameleon}), which share the most similarity among all datasets.
This is expected behavior when the autoencoder encounters out-of-distribution samples, as it maps them to the nearest known representation in the learned feature space~\cite{gadirov2021evaluation}.
We leave a deeper investigation of this behavior to future work.
\changedTo{}

\begin{table}[]
    \footnotesize
    \centering
    \caption{Further investigation of feature vector dimension.
    We highlighted the best prediction results in bold.
    }
     \label{tab:feature_select_2}
    
\begin{tabular}{@{}ccc@{}}
\toprule
\multirow{2}{*}{Model} & \multirow{2}{*}{$\mathrm{PSNR\uparrow}$/$\mathrm{T^{vol}_{infer}}\downarrow$} & $\mathrm{RMSE\uparrow}$ \\ \cmidrule(l){3-3} 
           &             & CUDA-raycaster       \\ \midrule
$128^3F2$  & 26.14/5.670 & \textbf{0.0049}          \\
$128^3F4$  & 26.82/5.700 & 0.0051 \\
$128^3F8$  & 27.03/5.776 & 0.0070         \\
$128^3F16$ & \textbf{27.27}/9.733 & 0.0071         \\ \bottomrule
\end{tabular}%
\end{table}

\textbf{(3) \acropred{} architecture selection.}
We now aim to determine a good balance between the final prediction accuracy and \acropred{}'s inference speed by considering the number of $\mathrm{C256}$ layers in \acropred{}.
We started without $\mathrm{C256}$ layers, and continued increasing the count until \acropred{} stopped improving (see \cref{tab:prednet_select_1}).
The overall trend is that as \acropred{} gets deeper, the prediction accuracy increases and reaches its peak. 
Then, the prediction accuracy of \acropred{} stopped increasing and started to decline due to overfitting.

For the CUDA raycaster, the optimal \acro{} architectures are $128^3$F$2\rightarrow12$C$256$.
For yt, the optimal \acro{} architectures are $64^3$F$32\rightarrow4$C$256$.
As expected, the model with the least number of layers has the fastest inference speed.
\begin{table}[t]
\footnotesize
\centering
  \caption{Comparison of RMSE of different \acropred{} architectures and inference time $\mathrm{T_{infer}^\mathrm{pred}}$.
Best predictions are in bold.}
  \label{tab:prednet_select_1}
  \begin{tabular}{@{}ccc@{}}
\toprule
\multirow{2}{*}{PredNet} & \multicolumn{2}{c}{$\mathrm{RMSE\downarrow}$/$\mathrm{T^{pred}_{infer}\downarrow}$} \\ \cmidrule(l){2-3} 
         & CUDA-raycaster        & yt                   \\ \midrule
$18C256$ & 0.0045/0.364          & -          \\
$16C256$ & 0.0046/0.324          & -          \\
$14C256$ & 0.0045/0.308          & -          \\
$12C256$ & \textbf{0.0045}/0.276 & -       \\
$10C256$ & 0.0046/0.244          & 3.917/0.244          \\
$8C256$  & 0.0046/0.219          & 3.925/0.222          \\
$6C256$  & 0.0047/0.190          & 3.949/0.190 \\
$4C256$  & 0.0048/0.163          & \textbf{3.823}/0.165          \\
$2C256$  & 0.0054/0.133          & 3.956/0.133          \\
$0C256$  & 0.0078/0.104          & 4.086/0.107          \\ \bottomrule
\end{tabular}%
\end{table}

\textbf{Final model selection.}
We have two ``optimal'' models in total (one for each  renderer). 
To select one model that performs well across all scenarios and datasets, we trained these two model architectures for both the CUDA raycaster and yt rendering scenarios on all datasets.
Then, we compared the prediction accuracy of each model with the best performance by calculating relative deviations.
The model with the lowest mean relative deviation was selected as our final model.
As shown in \cref{tab:final_select}, the optimal model is $128^3$F$2\rightarrow12$C$256$.

Overall, our model architecture selection shows that good performance can be achieved across a range of considered model configurations and scenarios.

\begin{table}[b]
    \footnotesize
    \centering
    \caption{Comparison of RMSE of different \acropred{} architectures and inference time $\mathrm{T_{infer}^{pred}}$.
    They are separated by ``/''.
    ``$\diamond$'' indicates that the results were from the baseline model. 
    RD denotes relative deviation from the best prediction and MRD is each model’s mean value of relative deviations. 
    We highlighted the best prediction accuracy and MRD in bold.}
    \label{tab:final_select}
    \begin{tabular}{@{}ccc@{}}
\toprule
\multirow{2}{*}{Renderer} & \multicolumn{2}{c}{$\mathrm{RMSE\downarrow}$/$\mathrm{T^{pred}_{infer}}\downarrow$/$\mathrm{RD}\downarrow$} \\ \cmidrule(l){2-3} 
                          & \multicolumn{1}{c|}{$\mathrm{128^3F2\rightarrow12C256}$}         & $\mathrm{64F32\rightarrow4C256}$         \\ \midrule
CUDA-raycaster            & \multicolumn{1}{c|}{$\diamond$\textbf{0.0045}/0.276/0\%}                                      & 0.0079/0.163/74.7\%                                   \\
yt                        & \multicolumn{1}{c|}{3.869/0.291/1.2\%}                                       & $\diamond$\textbf{3.823}/0.165/0\%                                    \\ \midrule
MRD                       & \multicolumn{1}{c|}{\textbf{0.6\%}}                                       & 37.4\%                          \\ \bottomrule
\end{tabular}%
\end{table}

\begin{figure}[t]
    \centering
    \subfloat[No scattering]{
    \begin{minipage}[b]{0.475\columnwidth}
        \centering
            \begin{tikzpicture}
                \begin{axis}[
                    xlabel={$\gamma$}, ylabel={RMSE}, 
                    xlabel style={font=\tiny, yshift=5pt}, xticklabel style={/pgf/number format/precision=3, font=\tiny}, 
                    xtick={0.2, 0.4, 0.6, 0.8, 1.0},
                    xticklabels={0.2, 0.4, 0.6, 0.8, 1.0},
                    ylabel style={yshift=-0.6cm, font=\tiny}, yticklabel style={/pgf/number format/precision=3, font=\tiny},  
                    width=1.15\columnwidth, height=4.5cm, xmin=0, legend pos=north east, legend style={fill opacity=0.0, draw=none, text opacity=1},]
                        \addplot[color=softred, line width=1pt,] 
                            table[x=ratio, y=rmse, col sep=comma]{files/num_sample_search/ns.txt};
                \end{axis}
            \end{tikzpicture}
    \end{minipage}
} 
\subfloat[Single scattering]{
    \begin{minipage}[b]{0.475\columnwidth}
        \centering
            \begin{tikzpicture}
                \begin{axis}[
                    xlabel={$\gamma$}, 
                    xlabel style={font=\tiny, yshift=5pt}, xticklabel style={/pgf/number format/precision=3, font=\tiny}, 
                    xtick={0.2, 0.4, 0.6, 0.8, 1.0},
                    xticklabels={0.2, 0.4, 0.6, 0.8, 1.0},
                    ylabel style={yshift=-0.1cm, font=\tiny}, yticklabel style={/pgf/number format/precision=3, font=\tiny},  
                    width=1.15\columnwidth, height=4.5cm, xmin=0, legend pos=north east, legend style={fill opacity=0.0, draw=none, text opacity=1},]
                        \addplot[color=softred, line width=1pt,] 
                            table[x=ratio, y=rmse, col sep=comma]{files/num_sample_search/ws.txt};
                \end{axis}
            \end{tikzpicture}
    \end{minipage}
}
    \caption{
    RMSE evaluated with varying numbers of training samples. 
    For each number of training samples, the experiment was conducted five times.
    The experiments demonstrated that the collected training samples are sufficient for training \acropred{}.
    }
   \label{fig:num_sample_search}
\end{figure}

\subsection{Study on the Number of Training Samples}\label{sec:TS}
We examined the effect of training sample size on prediction accuracy and demonstrated that collecting 100 rendering time measurements per volume is sufficient for training \acro{}.
For this study, we used the Nyx dataset with the CUDA-raycaster renderer.
\acro{}{} was evaluated under two scenarios: (1) without single-scattering, and (2) with single-scattering applied.
Their results are presented in \cref{fig:num_sample_search}.
For all scenarios, \acro{}{} was trained on a subset by selecting fewer samples from the full training set.
The ratio of samples used in comparison to the full set ($\gamma$) ranged from 10\% to 90\%.
Accordingly, $\gamma=100\%$ means the full set was employed for training \acropred{}.
Each experiment was conducted five times, with randomly selecting samples from the full training set at the same number of training samples.

For the scenarios with and without single-scattering, we observed that the RMSE decreased more slowly once the samplig ratio reached 0.6.
This suggests that increasing the number of training samples beyond this point yields diminishing returns in prediction accuracy, even in the more complex case where single-scattering is applied.
Thus, we demonstrate that collecting 100 samples per volume is sufficient to effectively train \acropred{}.

\subsection{Prediction Error Distribution Analysis}\label{sec:supp:error_analysis}
In this section, we analyze the prediction error distriburion of \acro{}.
\cref{fig:error_distribution} presents both the rendering time distributions (top) and squared prediction error distributions (bottom) across all datasets for the CUDA raycaster and yt renderer.
For the CUDA raycaster without single-scattering, all datasets exhibit tight, consistent distributions with minimal outliers, indicating stable and predictable GPU rendering performance. 
However, once single-scattering was applied (Nyx*), we observed significantly increased diversity in timing behavior, with a broader distribution and more frequent outliers extending to higher rendering times. 
This increased variability makes the prediction task considerably more challenging.

For yt, all three datasets display similar rendering time distributions, which aligns with the comparable prediction errors.
We observed that some outliers reach extremely high values (>400s), which can be attributed to our yt rendering being executed on a shared cluster where computational resources are occasionally overloaded due to concurrent jobs. 
We deliberately retain these outliers in our training data, as such performance variability represents realistic conditions when executing massive rendering tasks on supercomputers and HPC systems. 

The squared error distributions (bottom row) quantify ENTIRE's prediction accuracy. For CUDA raycaster, MAESTROeX, Nyx, and Castro achieve low squared errors with tight distributions. 
Nyx with single-scattering exhibits elevated squared errors.
While single-scattering shows numerically larger prediction errors due to more diverse timing distributions (similar to how CPU-based yt rendering shows wider distributions than GPU), ENTIRE maintains comparable prediction quality across all scenarios. The higher absolute RMSE for single-scattering reflects the increased variability in the underlying rendering times rather than degraded model performance.
For yt, all three datasets show similar error distributions.
Overall, the distribution of prediction errors aligns well with the rendering time behavior and the quantitative results.

\begin{figure}[t]
    \centering
    \input{figures/error_distribution}
    \caption{
    Prediction error distribution.
    (a) CUDA raycaster. (b) yt.
     $*$ indicates single-scattering was applied.
     Note the logarithmic scales.
    }
   \label{fig:error_distribution}
   \vspace{-12pt}
\end{figure}

\subsection{More Results for Use Case 1} \label{sec:supp:use_case_1}
In \cref{fig:control:curves}, we present all measured curves; the curve \( G \) that we use for final ray step size adaptation results from their median value at each ray step size. Notably, while the absolute rendering times vary, the overall trends remain consistent across all curves.
In our implementation, we access $G$ via a lookup table using linear interpolation (see \cref{fig:control:interpolation}).
\cref{fig:control_poses} shows three example poses for acquiring the look up table.

\begin{figure}[]
    \centering
        \subfloat[Sampled curves]{
    \label{fig:control:curves}
    \centering
      \begin{minipage}{0.5\columnwidth}
        \begin{tikzpicture}
          \begin{axis}[
          xlabel={Ray step size/$\delta^\mathrm{ref}$}, ylabel={Time/s},
          width=1.2\columnwidth, 
          xmin=0, xmax=31,
          xtick distance=2,
          xtick={5,10,15,20,25,30},
          xlabel style={yshift=0.25cm, font=\small}, 
          ylabel style={yshift=-0.7cm, font=\small}, 
          legend style={fill opacity=0.0, draw=none, text opacity=1, font=\tiny, legend columns=2},
          legend image post style={xscale=0.25},
          xticklabel style={font=\small},
          yticklabel style={font=\small},]
          \addplot[color=blue, line width=1pt]
          table[x=step, y=median, col sep=comma]{files/use_case_1/control_table.txt};
          \addlegendentry{$G(\delta)$}
          \addplot[color={rgb,255:red,192;green,214;blue,234}, line width=0.5pt, opacity=0.35]
          table[x=step, y=052_den_plt0000250_0, col sep=comma]{files/use_case_1/control_table.txt};
          \addlegendentry{(500, 0)}
          \addplot[color={rgb,255:red,166;green,199;blue,226}, line width=0.5pt, opacity=0.35]
          table[x=step, y=052_den_plt0000250_1, col sep=comma]{files/use_case_1/control_table.txt};
          \addlegendentry{(500, 1)}
          \addplot[color={rgb,255:red,134;green,183;blue,219}, line width=0.5pt, opacity=0.35]
          table[x=step, y=052_den_plt0000250_2, col sep=comma]{files/use_case_1/control_table.txt};
          \addlegendentry{(500, 2)}
          \addplot[color={rgb,255:red,208;green,144;blue,143}, line width=0.5pt, opacity=0.35]
          table[x=step, y=052_den_plt0001010_0, col sep=comma]{files/use_case_1/control_table.txt};
          \addlegendentry{(1000, 0)}
          \addplot[color={rgb,255:red,190;green,108;blue,109}, line width=0.5pt, opacity=0.35]
          table[x=step, y=052_den_plt0001010_1, col sep=comma]{files/use_case_1/control_table.txt};
          \addlegendentry{(1000, 1)}
          \addplot[color={rgb,255:red,170;green,58;blue,73}, line width=0.5pt, opacity=0.35]
          table[x=step, y=052_den_plt0001010_2, col sep=comma]{files/use_case_1/control_table.txt};
          \addlegendentry{(1000, 2)}
          \addplot[color={rgb,255:red,239;green,201;blue,155}, line width=0.5pt, opacity=0.35]
          table[x=step, y=053_den_plt0001270_0, col sep=comma]{files/use_case_1/control_table.txt};
          \addlegendentry{(1500, 0)}
          \addplot[color={rgb,255:red,232;green,181;blue,116}, line width=0.5pt, opacity=0.35]
          table[x=step, y=053_den_plt0001270_1, col sep=comma]{files/use_case_1/control_table.txt};
          \addlegendentry{(1500, 1)}
          \addplot[color={rgb,255:red,225;green,157;blue,73}, line width=0.5pt, opacity=0.35]
          table[x=step, y=053_den_plt0001270_2, col sep=comma]{files/use_case_1/control_table.txt};
          \addlegendentry{(1500, 2)}
          \end{axis}
         \end{tikzpicture} 
      \end{minipage}}
    \subfloat[Linear interpolation]{
    \label{fig:control:interpolation}
    \centering
      \begin{minipage}{0.5\columnwidth}
        \begin{tikzpicture}
          \node at (2.45,1.75) {\tiny{$\blacklozenge$} Theoretical point};
          \node at (2.45,1.25) {\tiny{$\medbullet$} Interpolation point};
          \begin{axis}[
          xlabel={Ray step size/$\delta^\mathrm{ref}$}, 
          width=1.2\columnwidth, 
          xtick distance=2,
          xlabel style={yshift=0.25cm}, 
          ylabel style={yshift=-0.75cm}, 
          legend style={fill opacity=0.0, draw=none, text opacity=1, font=\tiny}, 
          samples=100,
          domain=1:10,
          xmin=0, xmax=10,
          ymin=0, ymax=1,
          xticklabels={$\mathrm{n-3}$, $\mathrm{n-2}$,$\mathrm{n-1}$,$\mathrm{n}$,$\mathrm{n+1}$,$\mathrm{n+2}$,$\mathrm{n+3}$},
          xticklabel style={font=\small},
          ytick=\empty,
          xlabel style={font=\small},
          ylabel style={font=\small}]
          \addplot[color=navyblue, line width=0.5pt]
          {1/x};
          \addlegendentry{$G'(\delta)$}
          \addplot[color=magenta, line width=0.5pt] 
          coordinates {(1, 1/1) (2, 1/2)};
          \addplot[color=magenta, line width=0.5pt] 
          coordinates {(2, 1/2) (6, 1/6)};
          \addplot[color=magenta, line width=0.5pt] 
          coordinates {(6, 1/6) (8, 1/8)};
          \addlegendentry{$G(\delta)$}
          \addplot[black, line width=0.5pt, dashed] coordinates {(4, 0) (4, 1/3)};
          \addplot[black, line width=0.5pt, dashed] coordinates {(0, 1/3) (4, 1/3)};
          \addplot[black, line width=0.5pt, dashed] coordinates {(3, 0) (3, 1/3)};
          \addplot[only marks, mark=*, ] coordinates {(4, 1/3)};
          \addplot[only marks, mark=diamond*, ] coordinates {(3, 1/3)};
        \end{axis}
         \end{tikzpicture}       
      \end{minipage}}
\vspace{-1em}
    \caption{
    Our proposed ray step size adaption method.
    (a) Sampled $\delta$-$t^\mathrm{norm}$ curves.
    Time step and sample index pairs are denoted in the form of \textit{(time step, sample index)} in the legend.
    (b) Linear interpolation on $G(\delta)$; $G'$ depicts the theoretical GT curve.
    }
    \label{fig:control_methods}
    \vspace{-12pt}
\end{figure}

\begin{figure}[t]
    \centering
    \input{figures/control_poses.tex}
    \caption{
    Three poses for obtaining $G(\cdot)$.
    }
    \label{fig:control_poses}
\end{figure}

We developed an interactive widget for Use Case 1 based on the CUDA-raycaster.
The widget comprises two primary components: a panel for transfer function adjustment and a rendering window that supports interactive volume exploration.
The measured frame rate is displayed in real time for each rendered frame.

\begin{figure}[t]
    \centering
    \includegraphics[width=0.75\columnwidth]{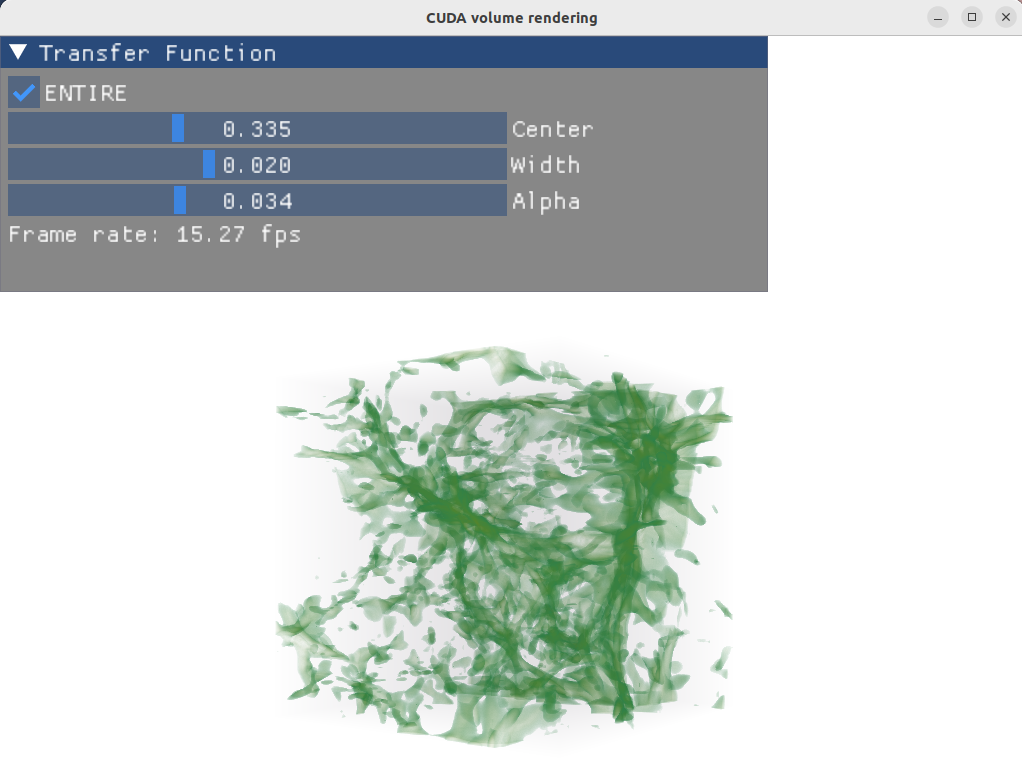}
    \caption{
    Screenshot of the widget for use case 1.
    The panel for transfer function adjustment and rendering time prediction model selection is placed at the top-left.
    The measured frame rate is displayed the bottom of the panel.
    The rendered volume is shown at the center of the window.
    }
   \label{fig:demo}
\vspace{-1em}
\end{figure}

\cref{fig:demo} presents the user interface designed for interactive volume exploration while maintaining a target frame rate.
The user can manipulate the volume in real time and adjust the transfer function using the slider in the control panel.
The resulting frame rate is displayed at the bottom of the interface, providing immediate feedback on performance.
A respective demonstration of user interaction with and without ray step size adjustment is included in the accompanying video.

\subsection{Further Discussion of Limitations and Future Work}


Here, we would like to extend upon our discussion of limitations and future work.

\textbf{Generalization.}
In future work, we aim to further investigate the implications of training a single combined model across these variations that is able to exploit commonalities across different variants.
This could significantly facilitate the deployment of \acro{} in practical scenarios such that no dedicated training is required but in a way the pre-trained weights can be used to achieve usable results already.
While with our data feature vector we can account for different volume data, similar means would be necessary to represent the other influence factors in such a generic model.
For example, for taking into account different compute hardware, we plan to capture relevant characteristics using micro-benchmarks; data resolution and ray step sizes could further be supplemented via additional model parameters.
Note, however, that this spans an increasingly high-dimensional parameter space that needs to be sampled adequately for efficient training (see the training efficiency discussion below).
Another, complementary effort regarding generalization involves investigating the effectiveness of \acro{} in predicting timings from other kinds of (rendering) methods in the context of computer graphics---i.e., for mesh-based data, potentially including global illumination approaches---or parallel volume rendering (see the discussion below on load balancing).

\textbf{Training Efficiency.}
In time series data such as used in this work, consecutive volumes often exhibit a high degree of similarity, leading to redundancy in our training dataset.
To address this, we plan to explore time step selection algorithms as a promising direction (e.g., \cite{frey2017flow,porter2019deep}).

\textbf{Architecture Improvements.}
We also aim to enhance \acrovol{}’s feature representation by leveraging recent advances in autoencoder architectures. 
For instance, Kaiming \textit{et al}.~\cite{he2022masked} demonstrated the effectiveness of an asymmetric autoencoder with a reduced decoder, trained on randomly masked images. 
Investigating similar architectures could further refine the volume encoding process and improve the overall predictive accuracy of \acro{}.

\end{document}